\documentclass[a4paper,fleqn,usenatbib]{mnras}

\usepackage{mathptmx}

\usepackage[T1]{fontenc}
\usepackage{ae,aecompl}

\usepackage{graphicx}	
\usepackage{amsmath,bm,color}	
\usepackage{amssymb}	
\DeclareMathOperator\erf{erf}
\newcommand{\bc}{}
\newcommand{\bcc}{}

\title[Induced planet formation in ``transition'' discs]{Dust traps as planetary birthsites: basics and vortex formation}

\author[Owen, J. E. \& Kollmeier, J. A.]{
James E. Owen,$^{1}$\thanks{E-mail: jowen@ias.edu}\thanks{Hubble Fellow}
and Juna A. Kollmeier,$^{1,2}$
\\
$^{1}$Institute for Advanced Study, Einstein Drive, Princeton NJ, 08540, USA\\
$^{2}$Observatories of the Carnegie Institution of Washington, 813 Santa Barbara Street, Pasadena, CA 91101, USA
}

\pubyear{2015}

\begin{document}
\label{firstpage}
\pagerange{\pageref{firstpage}--\pageref{lastpage}}
\maketitle

\begin{abstract}
We present a simple model for low-mass planet formation and subsequent evolution within ``transition'' discs.   We demonstrate quantitatively that the predicted and observed structure of such discs are prime birthsites of planets. Planet formation is likely to proceed through pebble accretion, should a planetary embryo ($M\gtrsim 10^{-4}$~M$_\oplus$) form. Efficient pebble accretion is likely to be unavoidable in transition disc dust traps, as the size of the dust particles required for pebble accretion are those which are most efficiently trapped in the transition disc dust trap. Rapid pebble accretion within the dust trap gives rise, not only to low-mass planets, but to a large accretion luminosity. This accretion luminosity is sufficient to heat the disc outside the gravitational influence of the planet and makes the disc locally baroclinic, and a source of vorticity.  Using numerical simulations we demonstrate that this source of vorticity can lead to the growth of a single large scale vortex in $\sim 100$ orbits, which is capable of trapping particles. Finally, we suggest an evolutionary cycle:  planet formation proceeds through pebble accretion, followed by vortex formation and particle trapping in the vortex quenching the planetary accretion and thus removing the vorticity source.  After the vortex is destroyed the process can begin anew.  This means transition discs should present with large scale vortices for a significant fraction of their lifetimes and remnant planets at large ~10AU radii should be a common outcome of this cycle. 

\end{abstract}

\begin{keywords}
accretion, accretion discs --- planets and satellites: formation --- planet-disc interactions --- protoplanetary discs
\end{keywords}



\section{Introduction}

Despite observational advances in characterising exoplanets and the environments in which they form (protoplanetary discs), a coherent picture that explains the origin and diversity of planets from their disky forebears remains elusive. We know planets must sculpt the discs in which they form, and indeed we see structures in observed protoplanetary discs; however, a lack of understanding means that using these observed structures to learn about the details planet formation is challenging.  

There is a handful of disc-systems which show IR colours and SEDs that are inconsistent with primordial discs: discs that are optically thick from small to large radii \citep[e.g.][]{Strom1989,Skrutskie1990}. Instead, these objects show inner regions that are heavily depleted in dust, while returning to primordial levels in the outer regions \citep[e.g.][]{Calvet2005,Espaillat2014,Owen2016}. Given these discs are partially cleared (in at least the dust) they have been termed ``transition'' discs. 

It is now known that transition discs do not represent a homogeneous class \citep[e.g.][]{OC12}. While many transition discs are believed to be protoplanetary discs caught in the act of clearing through photoevaporation \citep{Cieza2008,Merin2010,Owen2011}, a significant subset do not match any of the characteristics that would be associated with a protoplanetary disc in the throes of destruction \citep[e.g.][]{Kim2009,Espaillat2010,OC12}. {\bc Approximately half} of transition discs {\bc surveyed by \citet{OC12} had} large accretion rates ($\dot{M}\gtrsim10^{-9}$ M$_\odot$~yr$^{-1}$), large cleared dust cavities (out to $\gtrsim 10$~AU) and they {\bc were} often the  brightest of all class II discs at mm wavelengths \citep{Andrews2011,OC12,Ansdell2016}; for this reason they have been termed ``mm-bright transition discs" \citep{Owen2016}. 

The majority of mechanisms invoked to create a transition disc signature do so by removing dust from the inner regions, and preventing its resupply, by trapping dust outside some radius in a pressure enabled dust trap. Indeed this is true for the two most commonly invoked mechanisms: photoevaporation \citep[e.g.][]{Clarke2001,Alexander2007,Owen2011,Gorti2015} and gap formation by giant planets \citep[e.g.][]{Calvet2005,Rice2006,Zhu2011,Zhu2012,Owen2014}. 

Several of the mm-bright transition discs have been imaged at high resolution using {\it ALMA}, and in these high resolution images they show strong axisymmetric {\it and} non-axisymmetric emission features \citep[e.g.][]{Casassus2013,vanderMarel2013,Perez2014,vanderMarel2015,Andrews2016,Canovas2016} and see \citet{Casassus2016} for a recent review.  These features have been linked to signs of planet formation, although as yet there is no clear understanding of how to link these observations to theory. A common interpretation of the non-axisymmetric structures is that they caused by large-scale vortices \citep[e.g.][]{vanderMarel2013,Ataiee2013,LyraLin2013}. Since these vortices represent local pressure maxima that orbit the star with roughly the local Keplerian velocity they can efficiently trap dust particles \citep[e.g.][]{Meheut2012,LyraLin2013,Zhu2014}.  The strong dust density contrast that can result from dust particle trapping in vortices can lead to strong azimuthal surface brightness differences. The Rossby Wave Instability \citep[RWI][]{Lovelace1999,Li2000} provides a mechanism to generate  vortices in astrophysical discs that contain radial structure. Sharp radial features give rise to an extremum in potential vorticity that leads to vortex formation. Non-linear hydrodynamic simulations \citep[e.g.][]{Li2001} show that several small scale vortices grow from the linear instability before they begin to grow and merge resulting in one large scale vortex, which can trap particles \citep[e.g.][]{Lyra2009,Meheut2012,Zhu2014}.

Transition discs are believed to contain a cavity edge in the gas, and such drops in gas surface density have recently been observed in CO emission \citep{vanderMarel2015,vanderMarel2016}\footnote{\bc Note: CO is likely to be optically thick, and at this stage it is still difficult to directly transform CO structures into gas structures.}  . Such an axisymmetric gas structure is necessary to explain many of the observed features of transition discs, however, it remains to be seen whether the observed density drops are steep enough to trigger the RWI. While massive planets {\bc(of order Jupiter mass and higher)} inserted into discs in simulations are known to produce deep and sharp cavities, sharp enough to trigger the RWI, it is unclear whether this is likely to occur in reality. This is because massive planets are inserted into disc simulations instantaneously (or grown over several orbits), necessarily producing a transient phase with a very sharp cavity which is RWI unstable. 

In a realistic scenario, however, a planet accretes and grows on a time-scale comparable to the evolution of the protoplanetary disc itself. Viscosity present in the disc and, in fact, vortices created by any planet gap initially, can relax the gas to a much smoother distribution, which would be {\it stable} to the RWI over the long-term. This raises the intriguing question: if a transition disc cavity is stable to the RWI, or only unstable for a short period of time, why do transition discs, that are observable on time-scales $\gtrsim 10^4$ orbits, display non-axisymmetric structures generated by RWI vortices? This question is perhaps the biggest weakness of the RWI mechanism in explaining the observed transition disc structures \citep{Hammer2016}. 

{\bc The RWI can also be triggered in protoplanetary discs at the edge of the dead-zone \citep[e.g.][]{Lyra2009}, where the change in viscosity leads to a sharp change in the surface density \citep[e.g.][]{Gammie1996}. Recent hydrodynamic simulations have shown that large-scale vortices can also be produced this way \citep[e.g.][]{Regaly2012,Flock2015,Lyra2015,Ruge2016} through the RWI. {\bcc Recent work \citep{Ruge2016,Pinilla2016} has shown that the dead-zone model can produce rings and gaps in scattered-light and mm images, although it remains to be seen if the model can explain the large drops in dust-opacity in the inner disc sufficient to reproduce the SEDs of transition discs.} Finally, vortices can also be formed through baroclinic instabilities \citep[e.g.][]{Klahr2003,Lesur2010,Raettig2015}.}   

By trapping dust at some radius, while it continues to drift in from larger radii,  transition discs will have significant increases in the dust-to-gas ratio in the dust trap, from the standard ISM value of 0.01, to values that can approach unity \citep[e.g.][]{Pinilla2012}. Therefore, it has been suggested that these transition disc dust traps are likely to sites of increased planetesimal and planet formation. Indeed \citet{Lyra2008}, demonstrated that it is possible to get direct collapse to form planetary embryos in RWI generated vortices. 

One interesting advance in the theory of planet formation is the concept of pebble accretion \citep[e.g.][]{Ormel2010,Johansen2010,Lambrechts2012}. Dust particles that have gas-drag induced stopping times comparable to their orbital times can be rapidly accreted by planetary embryos. This is because gas pressure gradients induced by the planetary embryo's presence cause dust particles to be accreted if they approach within the proto-planet's Hill sphere, decreasing the growth time to $\sim 10^{4}$~years --- significantly faster than the standard planetesimal accretion times. Since pebble accretion is very sensitive to dust particle size --- it is only efficient for dust particles that are close to optimally coupled to the gas --- it can only work where there are large numbers of dust particles close to this size. The dust traps in transition discs {\it inevitably} provide a reservoir of these pebbles, as those dust particles that are most efficiently accreted are also those most efficiently trapped. 

Here we explore the possibility of low-mass planet formation in the pressure traps of transition discs, and argue that it is likely to naturally arise, while perhaps also explaining a variety of observational signatures that are now commonly associated with transition discs. We are agnostic about the mechanism that creates the transition disc itself, but argue that if low-mass planet formation begins in transition discs as we suggest it is: i) likely to be rapid and ii) if the disc is sufficiently massive, it is able to generate vortices that can lead to non-axisymmetric structures similar to those recently observed. 

We structure our work as follows: in Section~\ref{sec:mechanism} we discuss the physical picture and motivate the basic principles of our new mechanism. In Section~\ref{sec:sims} we present numerical simulations to look at the non-linear long term evolution. We discuss our results in Section~\ref{sec:results} and summarise in Section~\ref{sec:summary}.

\section{Physical Mechanism}
\label{sec:mechanism}

The characteristic feature of transition discs is the significant drop in opacity at small radii relative to a primordial disc with a return to primordial values at larger radii. This opacity drop is interpreted as a significant removal of dust (and hence opacity) close-in, while returning to standard values further out in the disc. The changeover from the optically-thin dust in the inner regions to the optically-thick outer regions is known to be sharp \citep[e.g.][]{Andrews2011}. Thus the standard explanation for such a dust distribution is that there is a ``dust-trap'', where a pressure maximum can overcome the gas accretion flow and trap dust particles in the vicinity of the pressure maximum due to gas-drag.  This is in contrast to a gradual change in the dust properties that might result from, for example, dust evolution \citep[e.g.][]{Dullemond2005,Birnstiel2012}.  

By generating an axisymmetric maximum in the gas surface density, which in turn creates a pressure maximum, dust particles will drift towards this pressure maximum where the dust particles feel no gas drag. The time-scale on which dust particles drift towards, and are trapped within the pressure trap, is determined by the particles' Stokes number, or the non-dimensional stopping time ($\tau_s$). Dust particles with $\tau_s$ {\bc greater than the viscous $\alpha$ can become trapped \citep[e.g.][]{Birnstiel2013}. However, it is those particles with $\tau_s$ near unity $\sim10^{-1}-10$ that are {\it rapidly} and {\it efficiently}} trapped in a pressure maximum on a time-scale $\lesssim100$ orbits. However particles with Stokes numbers far from unity are not strongly trapped, with diffusion and advection dominating for particles with very small Stokes numbers, meaning the dust closely follows the gas distribution. The size of a dust particle $s$ can be related to the Stokes number (in the Epstein drag limit - relevant for small dust particles in the outer disc) as:
\begin{equation}
s\approx2\,{\rm mm}\, \tau_s\left(\frac{\Sigma_g}{1~{\rm g~cm}^{-2}}\right)\left(\frac{\rho_d}{3~{\rm g}~{\rm cm}^{-3}}\right)^{-1}
\end{equation} 
where $\Sigma_g$ is the local gas surface density and $\rho_d$ is the dust density. Thus for typical parameters found in most transition discs, we find it is the mm to cm sized grains that are trapped in the pressure traps. This back of the envelope argument agrees with early mm observations of transition discs \citep[e.g.][]{Andrews2011}, which showed that the mm sized grain population was confined to narrow rings indicative of pressure trapping \citep{Pinilla2012}. 

Dust trapping can increase the surface density of particles with stopping times close to unity to very high values, of the order of 10~g~cm$^{-2}$ \citep[e.g.][]{Pinilla2012} in the most massive transition discs. Thus, the dust-to-gas ratio in the trap is significantly enhanced above the standard ISM value of 0.01. Such environments are ripe for the formation and growth of planetesimals and planetary embryos, either through coagulation or direct collapse through mechanisms such as the streaming instability \citep[e.g.][]{Youdin2005,Johansen2007}. The streaming instability is likely to be important in these dust traps since it is typically strongly triggered in environments with large dust-to-gas ratios \citep{Johansen2009}.  We suggest here that should a planetary embryo reach high enough mass such that it can undergoing pebble accretion, this planetary embryo will grow incredibly rapidly, as we now demonstrate. 

\subsection{Pebble Accretion in a Dust Trap}

Pebble accretion is a mechanism by which a planetary embryo can rapidly grow to significant mass \citep[e.g.][]{Lambrechts2012}. For dust particles that are coupled to the gas on time-scales comparable to the orbital time (i.e. $\tau_s\sim 1$), the embryo can accrete particles from impact parameters significantly larger than its physical radius or effective radius due to gravitational focussing. In fact, gas-drag enables the embryo to accrete particles from a radius out to its Hill sphere ($R_H=a(M_p/3M_*)^{1/3}$), with all particles with $\tau_s$ in the range 0.1-1, being accreted if they approach the embryo's Hill sphere \citep{Ormel2010,Lambrechts2012}.   

The rate at which embryos accrete through pebble accretion depends on whether it occurs in a largely planar or spherical manner \citep{Morbidelli2015,Bitsch2015}. In the planar case, the embryo accretes from the full height of the disc, whereas in the spherical case it just accretes from a fraction of the disc's height, and is therefore less efficient. The transition from spherical to planar accretion occurs approximately when the Hill sphere is larger than the scale height of the pebbles \citep{Morbidelli2015}. Or:
\begin{equation}
M_p \gtrsim 0.03~{\rm M}_\oplus \left(\frac{M_*}{1~{\rm M}_\odot}\right)\left(\frac{H/R}{0.1}\right)^3\left(\frac{\alpha}{10^{-3}}\right)^{3/2}
\end{equation}
As we shall see later, we are mainly concerned with accretion onto embryos with masses $\gtrsim 1$~M$_\oplus$ here, thus for simplicity, we assume pebble accretion will always take place in a 2D fashion with an accretion rate approximately given by:
\begin{equation}
\dot{M}_{\rm peb}\approx 2\Omega R_H^2 \Sigma_{\rm peb}
\end{equation}
As $R_H\propto M_p^{1/3}$, this means that at late times the planet mass will grow as $t^3$, provided it does not locally reduce the pebble surface density. With a typical growth time-scale of:
\begin{eqnarray}
t_{\rm acc}&\approx& 3\times 10^{3}\,\,{\rm yrs}\,\left(\frac{\Sigma_p}{3\,{\rm g~cm}^{-2}}\right)^{-1}\left(\frac{M_p}{5\,{\rm M}_\oplus}\right)^{1/3}\nonumber \\ &&\times \left(\frac{a}{20~{\rm AU}}\right)^{-1/2}\left(\frac{M_*}{1~{\rm M}_\odot}\right)^{1/6}
\end{eqnarray}
Such a time-scale is clearly very rapid, it could deplete the local reservoir of pebbles significantly, by turning them all into a low-mass planet and, as we shall argue in the next section, such a rapid accretion rate will have a significant thermal impact on the surrounding disc. 

Firstly, however, we will explore the time to deplete the dust trap using a simple model. Assuming that the dust trap is in a balance between turbulent diffusion and gas-drag and contains a mass in pebbles $M_{\rm peb}$, then the surface density of pebbles in the trap is given by $\Sigma_p=M_{\rm peb}/(2\pi R H_p)$, where $H_p$ is the radial width of the dust trap. Without considering the feedback on the gas $H_p$ could, in principle, be very thin compared to the disc's vertical scale height ($H$). Such a small $H_p$ would result in a dust-to-gas ratio well above unity, meaning dust drag becomes much less effective,  and as such the assumptions in deriving a thin $H_p$ break-down. Thus, we expect $H_p$ is likely to be fixed to the gas radial scale length when the dust-to-gas ratio approaches unity. As the dust-traps in transition discs are ordinarily expected to have dust-to-gas ratios well above the ISM value, for simplicity we set $H_p\sim H$ and leave a more detailed calculation to further work. Therefore, both the pebble reservoir and forming planet in a transition disc dust trap will evolve according to the following coupled equations:
\begin{eqnarray}
\frac{{\rm d}M_p}{{\rm d}t}&=&2\Omega a^2 \left(\frac{M_p}{3M_*}\right)^{2/3}\Sigma_p\label{eqn:planet_evolve}\\
\frac{{\rm d}\Sigma_p}{{\rm d}t}&=&\left(\frac{H}{R}\right)^{-1}\frac{\dot{M_p}}{2\pi a^2}\label{eqn:pebble_evolve}
\end{eqnarray}     
Equation~\ref{eqn:pebble_evolve} implies that at late times the pebble surface density drops exponentially, with a characteristic time-scale comparable to $t_{\rm acc}$ for a planet mass of roughly $M_{\rm peb}$. Therefore, we expect the planet mass to grow as $t^3$ until its mass becomes comparable to the mass originally in the pebble reservoir, at which point the reservoir is rapidly depleted in an exponential fashion. 

We are agnostic about the initial embryo mass since the strong temporal scaling of the growth makes our calculations largely insensitive to this value particularly at late times when the $t^3$ growth effectively erases the initial condition depleting all of the available mass into the planet.  Depending on the pebble surface density and separation, however, the absolute time-scale over which this occurs varies.  Nonetheless, this is always a rapid evolution occurring  in under $\sim 10^5$ years for a wide range of plausible surface densities and separations. 

\begin{figure*}
\centering
\includegraphics[width=\textwidth]{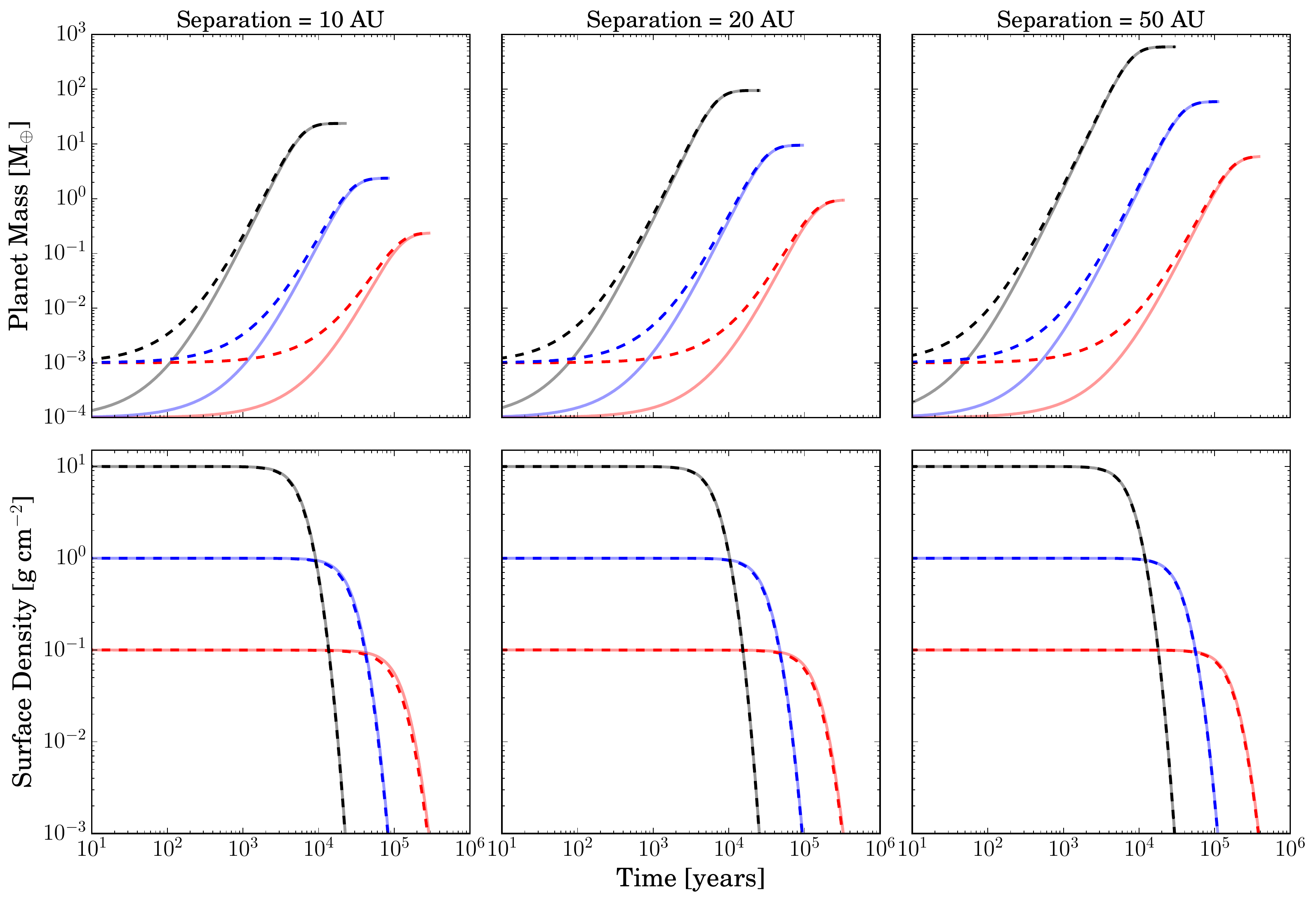}
\caption{The evolution of the planet mass (top row) and pebble surface density (bottom row) as a function of time. The solid lines show a starting embryo mass of $10^{-4}$~M$_\oplus$ and the dashed lines show a starting embryo mass of $10^{-3}$~M$_\oplus$. The line colour indicates the initial pebble surface density of 0.1 (red), 1.0 (blue) \& 10.0 (black) g~cm$^{-2}$. Each column show the calculations performed at separations of 10, 20 \& 50 AU.}\label{fig:evolve_res}
\end{figure*}

{\bc Pure coagulation could give rise to the initial embryo; bypassing the fragmentation barrier (which resides at sizes of order cm's) by the sweep-up mechanism (e.g.} Windmark 2012ab, Dr{\c a}{\.z}kowska et al. 2013{\bc), where larger particles can break through the fragmentation barrier by ``sweeping-up'' smaller ones.   We also highlight two other possible avenues for the formation of the initial embryos that could be present in transition disc dust traps}. First, direct gravitational collapse itself could proceed; while in a slightly different scenario (inside a RWI vortex) the dust surface densities are not too dissimilar \citet{Lyra2008} showed one could directly collapse to embryo sizes of order $10^{-2}$~M$_\oplus$ and above. Second, as we have hinted to above the dust traps are also prime sites for planetesimal and embryo formation through the streaming instability. \citet{Johansen2015} and \citet{Simon2016} showed through numerical simulations that one can grow embryos up to masses $\sim 10^{-4}$ M$_\oplus$. {\bc The streaming stability and coaugulation can also work together to promote embryo formation \citep{Drakowska2014}.} Certainly either of these scenarios or perhaps others could generate embryos large enough in mass to begin pebble accretion. Further, since the growth time drops as one approaches higher masses, accretion proceeds in a state that the first ``lucky'' embryo to start accreting will always be considerably more massive than all others. In the following calculations, we will assume the starting embryo masses are in the range $10^{-4}$-$10^{-3}$ M$_\oplus$, in accord with the values from the embryo formation models described above.  The gestation of these embryos into planets in our scenario is demonstrated in Figure~\ref{fig:evolve_res}, where we numerically evaluate Equations~\ref{eqn:planet_evolve} \& \ref{eqn:pebble_evolve} for parameters expected in transition discs.     
 Therefore, in this picture the embryos grow and deplete the local pebbles on a rapid time-scale that can reach planetary mass objects very quickly ($\sim 10^5$~years).   As seen in the Figure, this model naturally and rapidly produces super-Earth and Neptune-mass planets in these dust traps at large ($ \gtrsim 10$AU) radius from the host star. This formation channel could be one of the dominant modes of low-mass planet formation at large separations.

\subsection{Thermal Consequences}\label{sec:thermal}

Apart from providing the ideal site for the rapid formation of low-mass planets, and thereby exploiting the well-known virtues of pebble accretion, perhaps the most interesting and new consequence of this scenario is the thermal feedback from the rapid accretion. 

The energy resulting from the rapid accretion of pebbles will be liberated as radiation at the planet's surface, necessarily resulting in a large accretion luminosity. What we will argue is that if the low-mass planet can disrupt the thermal structure of the disc outside its Hill sphere (inside which the planet's gravity dominants) this can give rise to a source of vorticity strong enough to grow large scale vortices. In the outer regions of a protoplanetary disc we suspect it be optically thin to IR radiation \citep[e.g.][]{Chiang1997}. Thus, the temperature structure of the optically thin disc at a background temperature ($T_d$) surrounding a luminosity source is given by:
\begin{equation}
T=T_d\left[1+\left(\frac{R_T}{r}\right)^2\right]^{1/4}\label{eqn:temp_profile}
\end{equation}
where $R_T$ is the radius at which the black-body temperature due to just the irradiation from the planet is equal to the background disc temperature, and is given by:
\begin{equation}
R_T=\left(\frac{L}{16\pi\sigma T_d^4}\right)^{1/2}
\end{equation}
Thus assuming an accretion luminosity of $L=GM_p\dot{M}_{\rm peb}/R_p$ we can find the ratio of $R_T$ to the Hill radius as:
\begin{eqnarray}
\frac{R_T}{R_H}&=&\left(\frac{2GM_p\Omega\Sigma_{\rm peb}}{16\pi\sigma T_d^4R_p}\right)^{1/2}\\
&\approx & 5 \left(\frac{M_p}{5~{\rm M}_\oplus}\right)^{1/2}\left(\frac{R_p}{1.5~{\rm R}_\oplus}\right)^{-1/2}\left(\frac{\Omega}{3\times10^{-9}~{\rm s}^{-1}}\right)^{1/2}\nonumber \\&&\times \left(\frac{T_d}{30~{\rm K}}\right)^{-2}\left(\frac{\Sigma_{\rm peb}}{3~{\rm g~cm}^{-2}}\right)^{1/2}
\end{eqnarray}
Since the pressure trap can enhance $\Sigma_{\rm peb}$ to high values \citep[in some cases $\gtrsim 10$g~cm$^{-2}$, e.g.][]{Pinilla2012} the accreting planet will disrupt the temperature of the disc in regions significantly outside the Hill sphere of the planet. In fact in the above case, at the edge of the Hill sphere the temperature will be roughly three times the local disc temperature. Another relevant scale is $R_T/H$, which, for the above case, are roughly comparable.
\begin{eqnarray}
\frac{R_T}{H}&=&\left[\frac{2GM_p^{4/3}\Omega \Sigma_{\rm peb}}{16(3M_*)^{1/3} \pi\sigma T_d^4R_p(H/R)^2}\right]^{1/2}\\
&\approx&1.0 \left(\frac{H/R}{0.1}\right)^{-1}\left(\frac{M_p}{5~{\rm M}_\oplus}\right)^{2/3}\left(\frac{M_*}{1~{\rm M}_\odot}\right)^{-1/6}\nonumber\\&\times&\left(\frac{R_p}{1.5~{\rm R}_\oplus}\right)^{-1/2}\left(\frac{\Omega}{3\times10^{-9}~{\rm s}^{-1}}\right)^{1/2}\nonumber\\&\times&\left(\frac{T_d}{30~{\rm K}}\right)^{-2}\left(\frac{\Sigma_{\rm peb}}{3~{\rm g~cm}^{-2}}\right)^{1/2}
\end{eqnarray}
For distances from the planet $\gtrsim H$, the background Keplerian velocity, due to the shear, is supersonic with respect to the planet. If the temperature can disrupt the disc out to this radius it could cause wave breaking and significantly adjust the disc structure. However, given the weak scaling of $R_T/H$ on the parameters of the problem this is only likely to occur in the most extreme cases. 

\subsection{Radiative and advective time-scales}

Since the gas in the neighbourhood of the planet is orbiting at slightly different velocities to the planet due to the background shear, there is the obvious question of whether there is time for the gas parcels to be heated to the equilibrium temperature given by Equation~\ref{eqn:temp_profile}. One requires the advective time-scale to be shorter than the radiative timescale to be close to local radiative equilibrium. To first order the relative azimuthal velocity of a gas parcel with respect to the planet is given by:
\begin{equation}
v_{\rm rel}=\frac{3}{2}\Omega x_p
\end{equation}
where $x_p=|R-a|$ is the distance from the planet along the radial co-ordinate connecting star and planet. So the advective time $t_{\rm adv}\approx x_p/ v_{\rm rel}$ is independent of distance from the planet. Whereas the radiative time-scale is approximately:
\begin{equation}
t_{\rm rad}\approx\frac{4\pi r^2 k_bT_d}{\mu L \kappa}
\end{equation}
The ratio of radiative to advective time-scales is thus:
\begin{eqnarray}
\frac{t_{\rm rad}}{t_{\rm adv}}&=&\frac{3k_b\Omega}{2\mu \sigma T_d^3 \kappa} \left(\frac{r}{R_T}\right)^2\\
&\approx& 0.16 \left(\frac{r}{R_T}\right)^2 \left(\frac{\Omega}{3\times10^{-9}~{\rm s}^{-1}}\right)\\
&&\times\left(\frac{T_d}{30~{\rm K}}\right)^{-3}\left(\frac{\kappa}{0.1~{\rm cm^2~g^{-1}}}\right)^{-1}
\end{eqnarray}
Therefore, for nominal parameters we expect the surrounding gas to be in radiative equilibrium in the entire region which is actively heated by the planet, unless the opacity is low, but since we are in a dust trap with a high dust-to-gas ratio we would expect the opacity to be high, rather than low. {\bc For example, for a MRN particle distribution ($n(a)\propto a^{-3.5}$, as expected for a collisional cascade) that has a maximum particle size of 1~mm, \citet{dalessio2001} find an opacity of $\sim 0.1$~g~cm$^{-2}$ at FIR and mm wavelengths when the dust-to-gas ratio is $\sim 100$}.  We thus conclude that the gas will indeed reach the equilibrium temperature given by Equation~\ref{eqn:temp_profile} and the disc will have a ``hot-spot" associated with the rapid planet formation.

\subsection{Implications of hot-spot}

The fact the flow is no longer barotropic means no steady-state solution exists and, as we shall see in the numerical calculations, the temperature disturbance launches waves. However, we can gain a great deal of insight by considering how a temperature bump effects the flow structure along a closed  streamline in the disc\footnote{Note the following argument is {\it not} strictly a correct solution of the fluid equations in a rotating disc - we will use numerical simulations for that later - but is merely an illustrative calculation to gain insight.}.  The vertically integrated steady-state fluid equations along a streamline with path length, ${\rm d}\ell$, are given by:
\begin{eqnarray}
\frac{\partial}{\partial \ell}\left(\Sigma u \right)&=&0\label{eqn:cont2d}\\
u\frac{\partial u}{\partial \ell}+\frac{1}{\Sigma}\frac{\partial \mathcal{P}}{\partial \ell}&=&0\label{eqn:euler2d}
\end{eqnarray}
where $\mathcal{P}$ is the vertically integrated 2D pressure. Adopting a locally isothermal equation of state, such that the 2D pressure is given by $\mathcal{P}=\Sigma c_s^2$, where $c_s$ is the isothermal sound speed, we can combine Equations~\ref{eqn:cont2d} \& \ref{eqn:euler2d} to obtain a standard ``nozzle'' expression:
\begin{equation}
\left(\frac{u^2}{c_s^2}-1\right)\frac{\partial \log u}{\partial \ell}=-\frac{\partial \log c_s^2}{\partial \ell} \label{eqn:euler_mom}
\end{equation} 
As discussed above the relative velocity of gas parcels with respect to the planet does not become transonic until a distance of $\sim H$; therefore, the majority of the heated gas will have a significantly sub-sonic velocity with respect to the planet and temperature bump. This is crucial, as for the flow to adjust to the new temperature profile and reach dynamical equilibrium we require the flow time to be shorter than the sound crossing time, a condition obviously satisfied in the sub-sonic limit. 

Working in the sub-sonic limit with $r\lesssim H$, Equation~\ref{eqn:euler_mom} implies  that $\partial \log u/\partial \ell \approx \partial \log c_s^2/\partial \ell$. Since Equation~\ref{eqn:cont2d} relates the gradient of the surface density to the velocity, along with the equation of state we find:
\begin{eqnarray}
\frac{\partial \log \Sigma}{\partial \ell}&\approx&-\frac{\partial \log c_s^2}{\partial \ell}\\
\frac{\partial \log \mathcal{P}}{\partial \ell}&=&\frac{\partial \log \Sigma}{\partial \ell}+\frac{\partial \log c_s^2}{\partial \ell}\approx 0
\end{eqnarray}
Namely, the 2D pressure remains approximately constant as the gas parcel passes through the temperature bump, while the surface density drops, inversely tracking the temperature profile. This is an important result as it means the flow in the vicinity of the planet has become {\it baroclinic}. The implications of this fact become clear if we inspect the inviscid vortencity equation for a 2D disc:
\begin{equation}
\frac{D}{Dt}\left(\frac{\bm{\omega}}{\Sigma}\right)=\frac{\bm{\nabla}\Sigma\times\bm{\nabla} \mathcal{P}}{\Sigma^3}
\end{equation}
where $\bm{\omega}$ is the vorticity, which is only non-zero in the $\bm{\hat{z}}$ direction in our 2D disc. Therefore, if the flow was completely barotropic then vortencity ($\bm{\omega}/\Sigma$) would be conserved. However, along a gas parcel orbit, in the vicinity of the planet, $\bm{\nabla}\Sigma\times\bm{\nabla}\mathcal{P}\ne\bm{0}$ and the hot spot can source vortencity, allowing vortex structures to grow.


Therefore, in this work we present a mechanism for the generation and maintenance of vortices in astrophysical accretion discs. We discuss the picture specifically within the case of transition discs as this is likely to be observationally relevant; however, we note it should occur more generally, whenever $R_T\gtrsim R_H$ for a forming planet, or for cases where there is a luminosity point source in an astrophysical accretion disc.\footnote{Our framework describes the general phenomenon of hot-spot induced anti-cyclones in astrophysical accretion discs.  Here, the vortex is induced by planet-formation in a transition disc but in other systems, we anticipate a different heat source and different disc type.}

\section{Numerical Calculations}
\label{sec:sims}
In order to investigate the long term evolution of a disc with a planet induced temperature bump and to see if we can grow large scale vortices we must perform numerical simulations. We work in 2D and use the {\sc fargo} code \citep{Masset2000} and adopt a locally isothermal equation of state ($\mathcal{P}=\Sigma c_s^2$), where we have modified the code to include a 2D locally isothermal temperature distribution. We consider the evolution of a low-mass planet that induces a temperature bump within a transition disc like gas structure. {\bc To construct the simplest possible model, in order to isolate and investigate the physics of our new process, we neglect planet migration, dust-gas coupling, dust evolution, and cooling, all of which should be addressed in future work.}

\subsection{Setup}

As the starting point for our background transition disc gas structure we take a standard protoplanetary disc model with a surface density and passively heated temperature profile:
\begin{eqnarray}
\Sigma_b&=&\Sigma_1\left(\frac{R}{R_1}\right)^{-1}\\
T_b&=&T_1\left(\frac{R}{R_1}\right)^{-1/2}
\end{eqnarray}
where the normalising temperature at $R_1$ is chosen such that $H/R=0.1$ at $R_1$. In order to insert a gas cavity into the disc to mimic the possible gas structure that is likely present in a transition disc, we modify our background surface density structure ($\Sigma_b$) to the form:
\begin{equation}
\Sigma=\frac{\Sigma_b}{2}\left[1+\erf\left(\frac{R-R_0}{\sqrt{2}\sigma}\right)+\epsilon\right]
\end{equation}
where $R_0$ is a parameter that can be used to set the location of the peak of the surface density distribution, which we take to occur at $R_1$, such that $R_0\approx0.75 R_1$, and $\epsilon$ is a small number chosen to prevent the surface density from becoming zero at small radius {\bc in the initial surface density distribution}, which we set to  $10^{-3}$. The parameter $\sigma$ controls the smoothness over which the surface density declines. If $\sigma$ is too small then the disc will be unstable to the standard RWI and will naturally form vortices. This is obviously not what we want to investigate here, and we choose a value of $\sigma=0.3R_1$. We note we are not appealing to any physical model here (e.g. photoevaporation or giant planets), rather we want to remain agnostic and study a disc that has the general features one might expect in a transition disc, while not starting from a disc structure that is RWI unstable.

The angular velocity is initially set to the Keplerian value, {\bc suitably adjusted for the pressure gradient to maintain radial dynamical balance, using the numerical procedure suplied in {\sc fargo}.} We also adopt wave damping boundary conditions. We have checked that our initial surface density and temperature profile is stable by evolving it forward for 500 orbits at $R_1$. In order to isolate the physics of our new mechanism we ignore disc self-gravity and the indirect potential at this stage, meaning our choice of surface density normalisation is arbitrary and we scale all results in terms of $\Sigma_1$. 

We insert a planet in the disc with mass ratio $q=M_p/M_*$ on a fixed circular orbit at $R_1$, and the local temperature is modified such that it has the profile:
\begin{equation}
T=T_b\left[1+\left(\frac{R_L^2}{r^2+s^2}\right)\exp\left(\frac{-r^2}{2(fR_L)^2}\right)\right]^{1/4}\label{eqn:T_numerics}
\end{equation}
where $r$ is the distance from the centre of the planet. The temperature expression (Equation~\ref{eqn:T_numerics}) contains two additional terms not present in Equation~\ref{eqn:temp_profile}. The factor $s$ is a smoothing distance so that the temperature profile does not diverge on the grid and the exponential term provides a cut-off in the temperature profile at radii so far from the planet that gas parcels do not spend sufficient time in the planet's vicinity such that they are heated, with the cut-off radius occurring at a radius of $fR_T$. While we shall see the choice of $f$ does affect our results, $s$ doesn't as the smoothing occurs on the grid scale, well inside the planet's Hill sphere. 

\begin{table*}
\centering
\begin{tabular}{l|ccccccc}
Simulation & $N_R\times N_\phi$ & $q$ & $R_T/R_1$ & $R_T/R_H$ & $f$ & $s$ & $\alpha$ \\
\hline
Ctrl1 & $768\times1024$ & 0 & --- & --- & ---& ---& $1\times10^{-6}$ \\
Ctrl2 & $768\times1024$ & $1.5\times10^{-5}$ & 0 & 0 & ---& ---& $1\times10^{-6}$ \\
Standard & $1024 \times 1408$ & $1.5\times10^{-5}$ & 0.06 & 3.5 & $\infty$ & $9\times10^{-3}$ & $1\times10^{-6}$ \\
CutOff & $1024 \times 1408$ & $1.5\times10^{-5}$ & 0.06 & 3.5 & 1 & $9\times10^{-3}$ & $1\times10^{-6}$ \\
LowLum & $1532 \times 2056$ & $4.5\times10^{-5}$ & 0.01 & 0.4 & 1 & $5\times10^{-3}$ & $1\times10^{-6}$ \\
Viscous1 & $1024 \times 1408$ & $1.5\times10^{-5}$ & 0.06 & 3.5 & $\infty$ & $9\times10^{-3}$ & $1\times10^{-4}$ \\
Viscous2 & $1024 \times 1408$ & $1.5\times10^{-5}$ & 0.06 & 3.5 & $\infty$ & $9\times10^{-3}$ & $5\times10^{-4}$ \\
Viscous3 & $1024 \times 1408$ & $1.5\times10^{-5}$ & 0.06 & 3.5 & $\infty$ & $9\times10^{-3}$ & $1\times10^{-3}$ \\
\hline
\end{tabular}
\caption{Description of simulation parameters run. The second column lists the simulation resolution, columns 3--7 list physical and numerical parameters described above and the final column lists the $\alpha$ viscosity included.}\label{tab:sim_params}
\end{table*}

All simulations are performed on a polar grid with logarithmic-spaced cells in the radial direction and uniform-spaced cells in the azimuthal direction. This choice keeps the ratio of the cell side lengths approximately square over the entire domain. The radial grid stretches from 0.1$R_1$ to 10$R_1$, and we consider the full $2\pi$ in the azimuthal direction. We use the {\sc fargo} transport algorithm to speed up time-stepping, and work in a frame that is co-rotating with the planet. An explicit viscosity is included using an $\alpha$ viscosity law. The parameters for the simulation runs are described in Table~\ref{tab:sim_params}. We pick our temperature smoothing length $s$ to correspond to approximately two grid cells and we also smooth the planet's potential with a length scale of 0.5 the planet's Hill radius using the {\sc RocheSmoothing} facility implemented in {\sc fargo}. The planets are inserted at zero time with their full masses whereas the temperature hot spot's size $R_T$ is grown linearly over the first 50 orbits to the desired value.

\subsection{Simulation Results}

\begin{figure*}
\centering
\includegraphics[width=\textwidth]{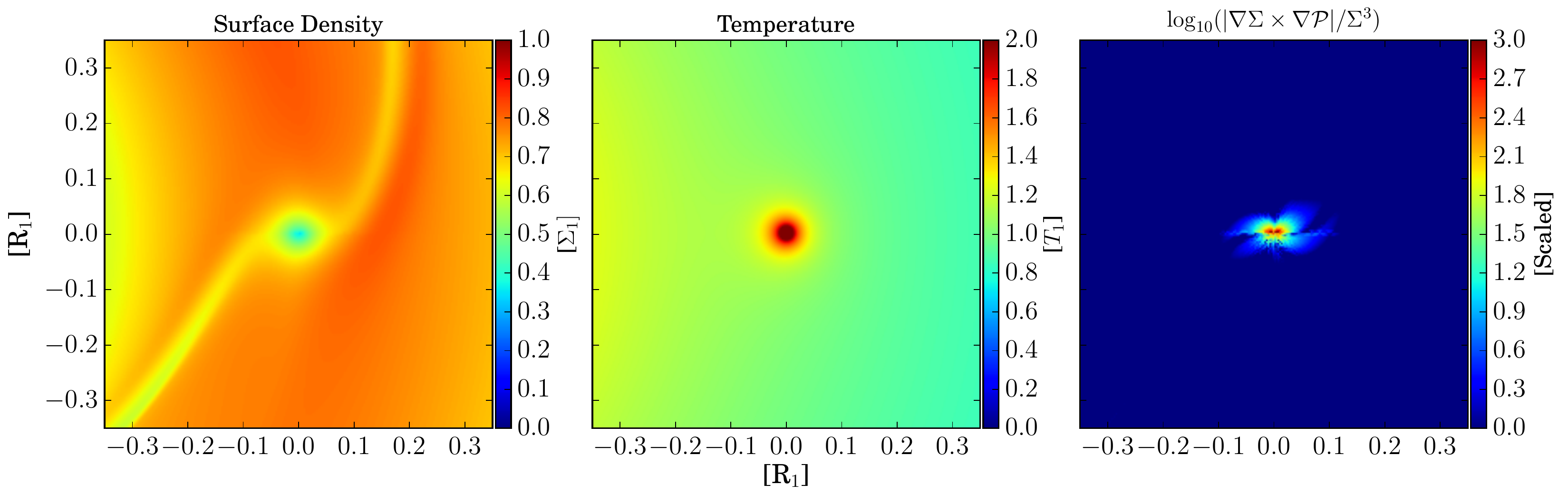}
\caption{Zoom-in on the planet showing the surface density (left), temperature (centre) and magnitude of the baroclinic vector (right). The simulation shown is the Standard case with $R_T/R_H=3.5$ and $f=\infty$, and is shown after 50 orbits once the hot spot has grown to its full size, but before large scale vortex formation has occurred. {\bcc The units $\Sigma_1$ and $T_1$ are defined in Equations 24 \& 25 respectively as the surface density and temperature at $R_1$ of the backgroud, power-law disc profile.}}\label{fig:baroclinic}
\end{figure*}

We evolved all of our simulations for 300 orbits to asses the impact of a hot spot around a planet in a transition disc-like gas structure. As discussed in Section~\ref{sec:thermal} the hot spot results in a surface density decrease that scales with the temperature increase. The resulting flow properties in the vicinity of the planet for our Standard case are shown in Figure~\ref{fig:baroclinic}. The properties are shown after 50 orbits; once the hot spot has grown to its full size, but before the resulting dynamics begin to dominate the flow. 

Figure~\ref{fig:baroclinic} shows the surface density (left), temperature (centre) and baroclinic vector (right) for a zoom-in on the planet in the Standard run. We see a significant reduction in the surface density in the vicinity of the planet that mirrors the temperature increase as expected. We also note that the temperature hot-spot also launches waves that dominate over the density waves that are purely driven by the planet's gravity. This surface density drop and temperature increase correspondingly results in a significant baroclinic term that is 2-3 orders of magnitude larger than those arising from the planetary wakes. This source of vorticity extends well outside the planet's Hill sphere (which has a radius of $\sim 0.17R_1$) and thus gas parcel orbits intersect this region every orbit, allowing for the growth of vorticity.  

\begin{figure*}
\centering
\includegraphics[width=\textwidth]{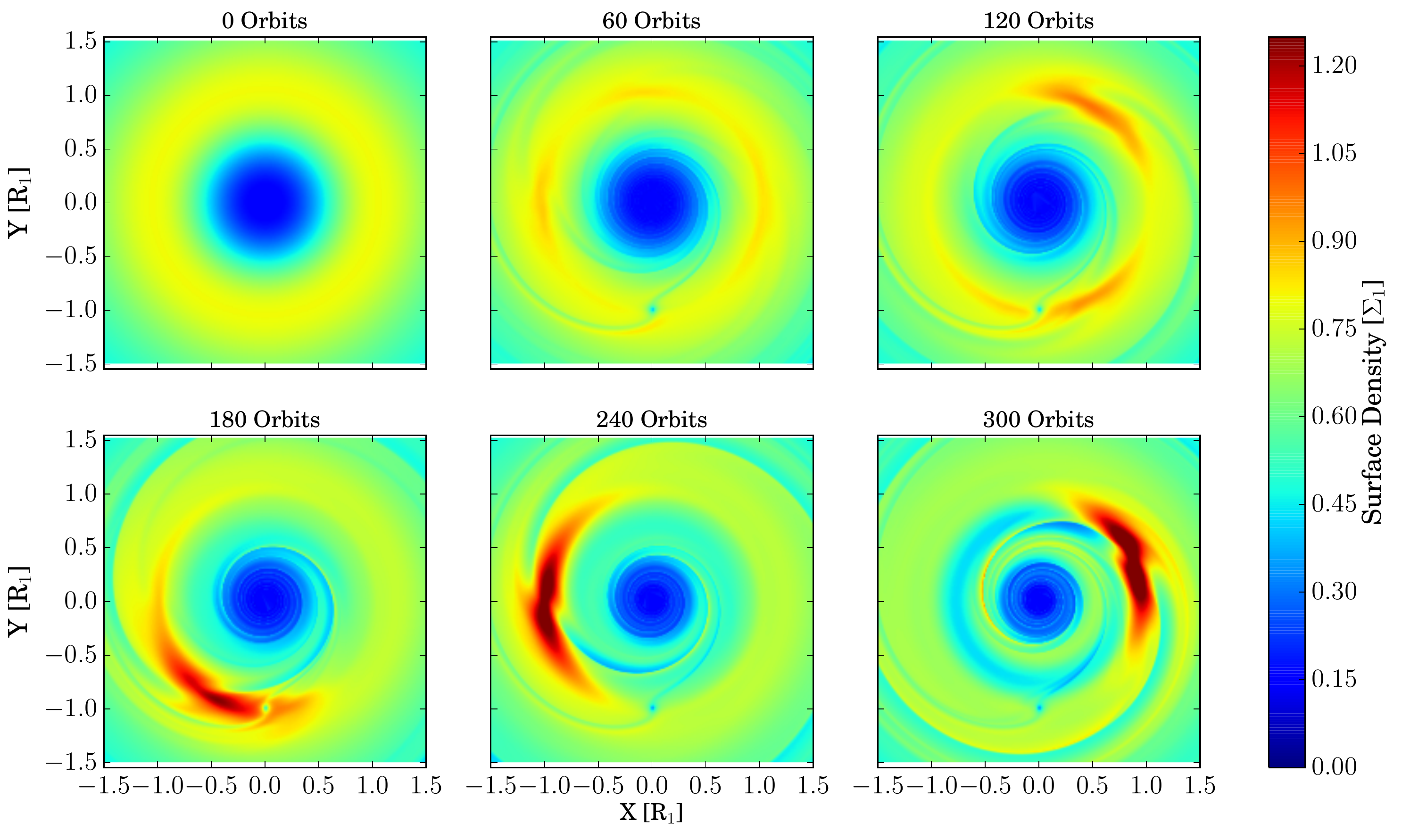}
\caption{Snapshots of the surface density distribution every 60 planetary orbits for the Standard run with $M_p/M_*=1.5\times10^{-5}$, $R_T/R_H=3.5$ and very low viscosity of $\alpha=10^{-6}$. The planet is located at a position $[0,-1]$.}\label{fig:Standard_Sevol}
\end{figure*}

\begin{figure*}
\centering
\includegraphics[width=\textwidth]{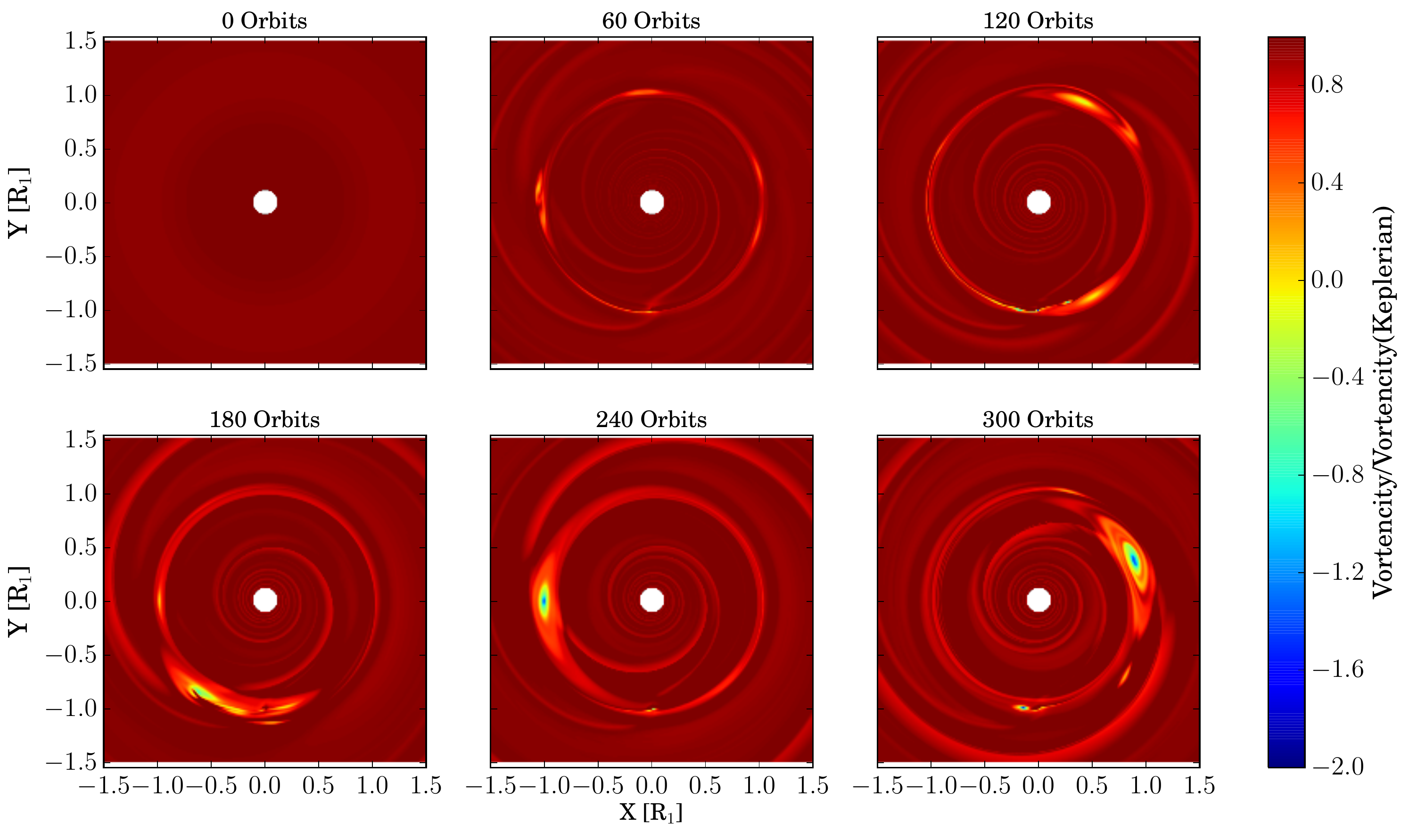}
\caption{Snapshots of the vortensity (scaled to the value of the disc with a Keplerian velocity profile) shown every 60 orbits for the Standard run. The planet is located at a position $[0,-1]$.}\label{fig:vortensity_evol}
\end{figure*}

The evolution of the surface density and vortencity of the Standard run are shown every 60 orbits in Figures~\ref{fig:Standard_Sevol} \& \ref{fig:vortensity_evol} respectively. These show that the hot spot rapidly produces a number of small scale vortices that grow (thorough interactions with the hot spot) and merge.  After approximately 150 orbits (100 orbits since the hot spot reached full size), the disc contains a significant large scale anti-cyclonic vortex that has grown to the maximal width of $\sim 2H$. The azimuthal surface density enchantment has reached roughly a factor of two over the original transition disc gas structure. Over the continued evolution, the vortex grows in strength and after 300 orbits the vortex has begun to migrate. Once the vortices have merged to result in a single large scale vortex it has pattern speed with respect to the planet in the range $|\Omega_{\rm pattern}/\Omega_p-1|\approx0.01-0.05$, close enough to Keplerian such that it should efficiently trap particles \citep{Ataiee2013}.

\begin{figure*}
\centering
\includegraphics[width=\textwidth]{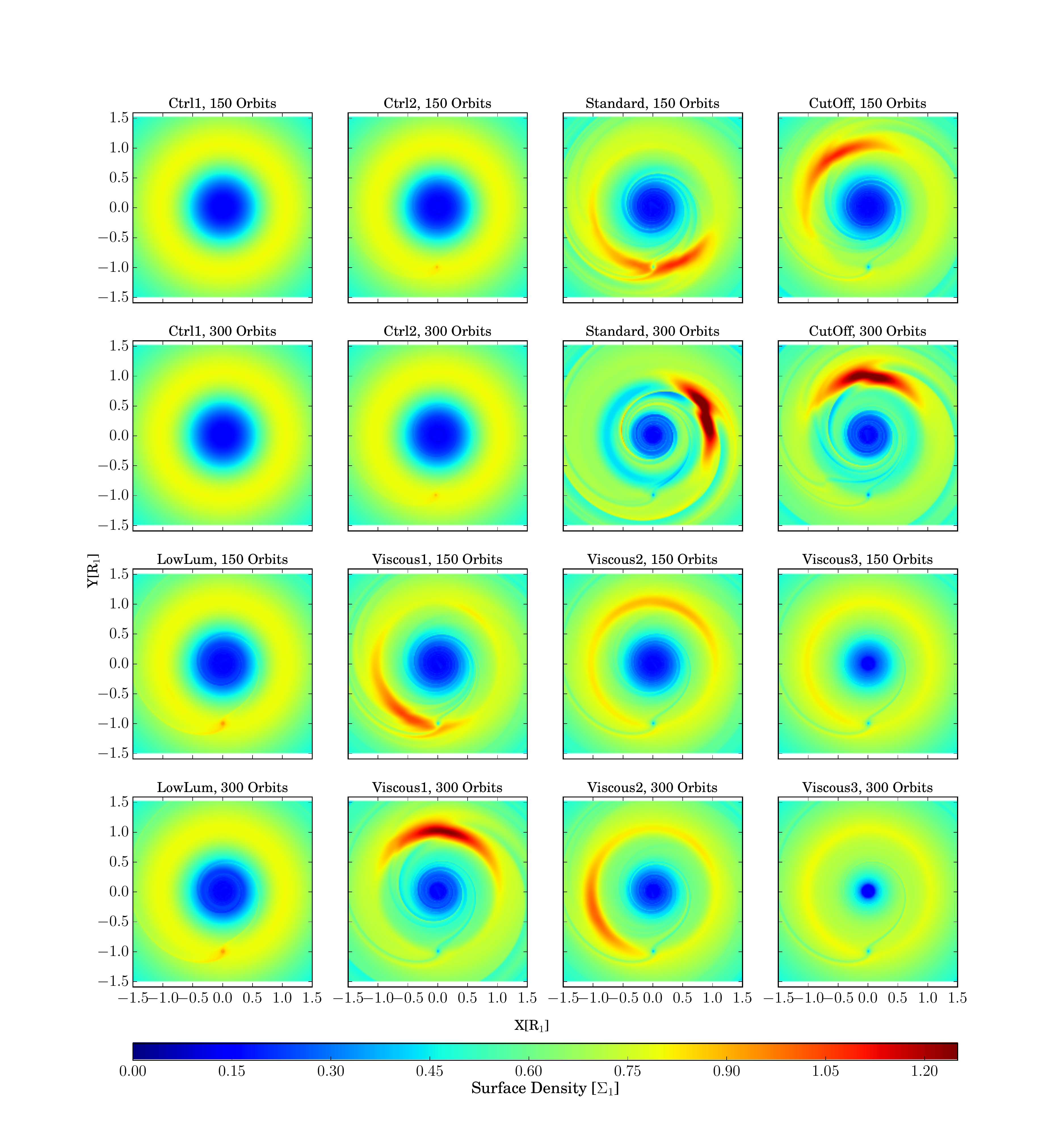} 
\caption{Snapshots of the surface density distribution for each simulation shown after 150 and 300 orbits. The planet is located at a position $[0,-1]$. The simulation parameters are indicated in Table~\ref{tab:sim_params}.}\label{fig:evol_all}
\end{figure*}

The surface density snapshots at 150 and 300 orbits are shown for all our simulations in Figure~\ref{fig:evol_all}. In the three cases where there is no significant source of baroclinicity -- Ctrl1, Ctrl2, LowLum -- no large scale vortex forms as expected. In Ctrl1, where we have no planet, as in all simulations we have deliberately selected a starting gas profile that is not steep enough to be RWI unstable, and as such the disc profile evolves unchanged for many hundreds of orbits. In Ctrl2, the planet has a mass of $M_p/M_*=1\times10^{-5}$ (which is well below the thermal gap opening mass), while the planet launches standard density waves as expected, it does not effect the large scale dynamics of the disc. In simulation LowLum, we have $R_T<R_H$ and the thermal impact of the planet is contained well within the Hill sphere of the planet. As such no source of significant baroclinicty  intersects with circular gas parcel orbits and, as expected, no large scale vortex is generated.

In the CutOff simulation, where we mimic the effect of the finite thermal inertia of the gas with an exponential cut-off to the temperature profile, the weakened source of vorticity results in a slightly weaker vortex; however, the general differences are small and a large scale vortex still forms after $\sim 100$ orbits. 

\subsubsection{Effect of Viscosity}
Viscosity is known to effect vortex growth and survival in protoplanetary discs and indeed this is what our simulations with significant viscosity indicate. We find that with $\alpha=10^{-4}$ in the Viscosity1 run, the vortex growth is barely suppressed compared to the Standard run with negligible viscosity. In Viscosity2 with $\alpha=5\times10^{-4}$, large scale vortex formation still does occur but it is suppressed with an azimuthal surface density contrast of $\sim 1.3$ compared with $\sim 2$ in the Standard run. 

In the most viscous run - Viscosity3 - with $\alpha=10^{-3}$, we find that there is a weak vortex present at 150 orbits, with an azimuthal density contrast of $\sim 1.05-1.1$; however, it does not grow to become a strong large scale vortex and after 300 orbits the disc has returned to being close to axisymmetric. At this high viscosity the gas surface density bump is also significantly weakened due to standard radial viscous transport, allowing the small vortices initially generated to migrate into the inner regions of the disc. Thus, while the higher viscosity certainly makes large scale vortex formation more difficult, it is unclear whether if some process existed to maintain the gas profile during the simulation (e.g. photoevaporation, a dead-zone or a massive planet) a large scale vortex would have indeed formed.



\section{Discussion}
\label{sec:results}

We have shown that ``transition'' discs are prime sites for the growth of low-mass planets by pebble accretion. This planet formation scenario has two important -- perhaps mutually exclusive -- implications. Firstly, if the planet were able to accrete pebbles at the standard rate for the entire time planet formation is occurring, then it could quickly deplete the local pebble reservoir in a very rapid $\lesssim 10^{5}$~year time-scale. Secondly, if the accretion rate results in a large accretion luminosity, such that it can heat the disc material outside its Hill radius, then it can lead to large scale vortex formation. The first implication begs an interesting question: if the pebble depletion time-scale is so fast $\lesssim 10^{5}$~years, how is it we see a number of transition discs with large mm-fluxes (and hence large reservoirs of pebbles) that almost certainly have lifetimes $\gtrsim 10^{5}$ years?

We hypothesise the answer to this question lies in the second consequence. The observed, probably long lived, ``transition'' discs that have large mm-fluxes \citep[e.g.][]{Andrews2011,OC12,Owen2016}, are those discs which are likely to have the highest pebble accretion rates, and as such the discs most prone to vortex formation. Vortex formation, results in an azimuthal pressure trap that can very efficiently trap pebbles \citep[e.g.][]{Meheut2012,Ataiee2013,Birnstiel2013,Zhu2014}. After all, the same dust particles that are likely to be trapped in the dust trap, are those which will accrete onto the planet, and are also those most likely to be trapped in the vortex. Since the vortex can migrate slightly, and does not necessarily have a pattern speed that is exactly Keplerian (as seen in our simulations), then the pebbles that can accrete on to the planet are trapped in the vortex and can only accrete onto the planet for a small fraction of time, or none if the vortex and planet have migrated apart. 

\begin{figure}
\centering
\includegraphics[width=\columnwidth]{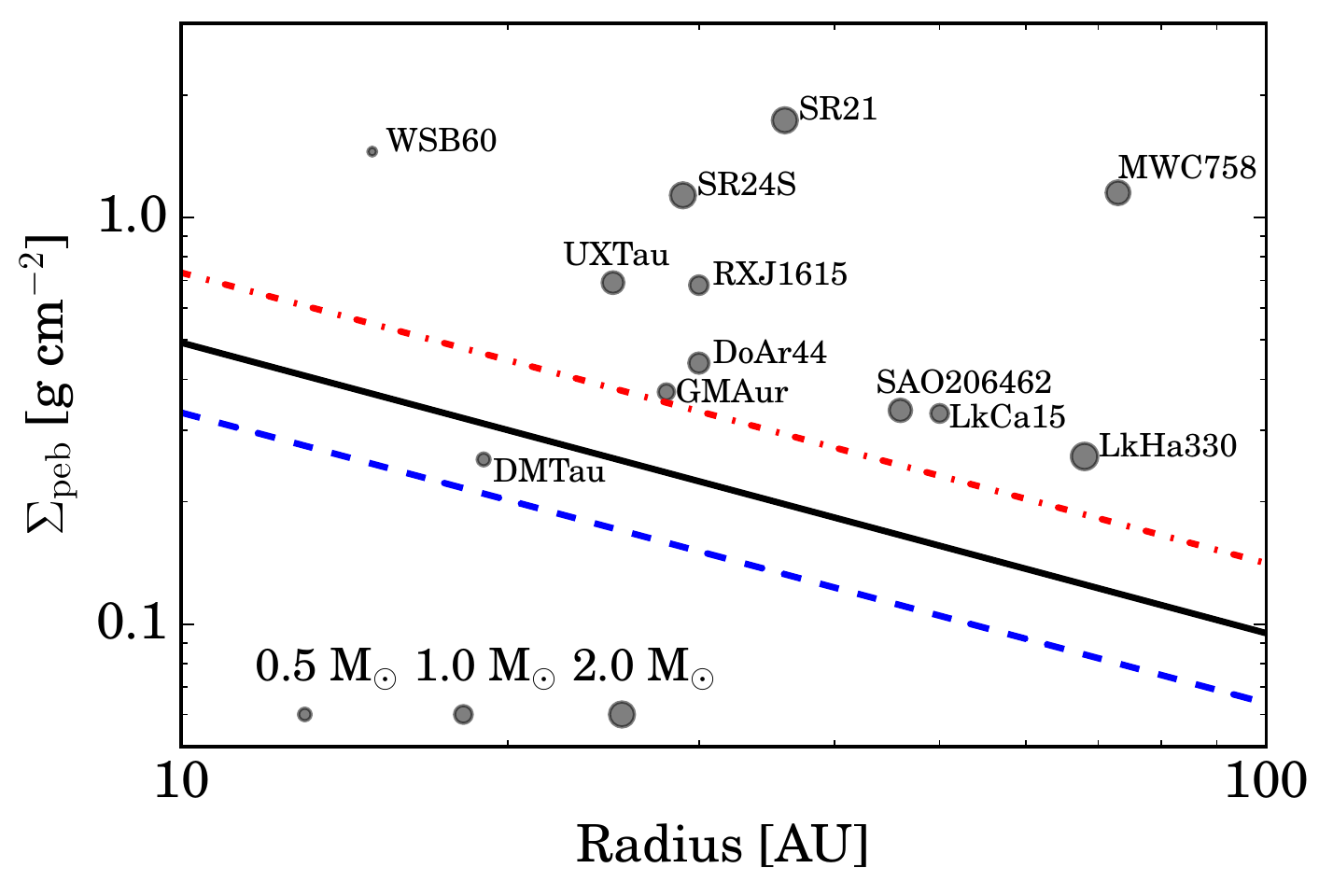}
\caption{The peak surface density of mm particles determined from mm imaging for well known mm-bright ``transition'' discs taken from \citet{Andrews2011} shown as points. The point sizes indicate the stellar mass. The lines show the minimum pebble surface density for vortex formation to be possible indicated by Equation~\ref{eqn:sigma_crit2}, the lines show different stellar masses: 0.5 (dot-dashed), 1.0 (solid) \& 2.0 M$_\odot$ (dashed). {\bc The labels show the individual source names.} }\label{fig:Sigma_crit_compare}
\end{figure}

Therefore, vortex formation and subsequent particle trapping may be the only way to ensure that these mm-bright transition discs remain long-lived. In fact, we can calculate the critical pebble surface density for a transition disc to form a vortex, and in principle remain long lived. We do this by assuming this threshold occurs when $R_T/R_H>1$. Since the Hill radius scales with planet mass as $M_p^{1/3}$ and $R_T$ scales as $M_p^{17/24}$  (where the solid planet mass-radius relationship can be approximated as $M_p=A_{\rm MR}R_p^4$ in the range 1-10 M$_\oplus$ using the profiles from \citealt{Fortney2007}). Thus, the limiting case when a disc can cross the $R_T/R_H>1$ threshold will be at the highest mass the planet can possibly reach. Therefore, if we approximate the maximum mass a pebble accreting planet can reach as $M_p\approx\Sigma_{\rm peb}/2\pi aH_p$ (i.e. the mass it would reach if it had accreted all the pebbles) and assume it is accreting from a reservoir of pebbles with surface density $\Sigma_{\rm peb}$, then we estimate the critical pebble surface density for vortex formation to occur before the entire reservoir is sequestered into a planet as:
\begin{equation}
\Sigma^{\rm crit}_{\rm peb}=\left[\frac{16\pi\sigma T_d^4A_{\rm MR}^{1/4}}{2G\Omega\left(2\pi a H_p\right)^{3/4}}\right]^{4/7}\label{eqn:sigma_crit}
\end{equation}
Given a $T\propto R^{-1/2}$ temperature profile for the passively heated, flared disc \citep[e.g.][]{Kenyon1987}, Equation~\ref{eqn:sigma_crit} only depends on a few parameters such that:
\begin{equation}
\Sigma^{\rm crit}_{\rm peb}\approx 0.3\, {\rm g\, cm^{-2}} \left(\frac{H_p}{H}\right)^{-3/7}\left(\frac{a}{20\,{\rm AU}}\right)^{-5/7}\left(\frac{M_*}{1\,{\rm M}_\odot}\right)^{-4/7}\label{eqn:sigma_crit2}
\end{equation}
where we have left $H_p/H$ as a free parameter here, but we suspect it to close to unity as discussed above. This critical surface density threshold can be compared to many of the well known mm bright transition discs. This comparison is shown in Figure~\ref{fig:Sigma_crit_compare}, where we plot the peak surface density at the edge of the cavity determined from mm imaging by \citet{Andrews2011} (using their model fits), compared to the result from Equation~\ref{eqn:sigma_crit2}. {\bc This is done by taking the surface density models provided in \citet{Andrews2011}, which are obtained from fits to the mm image and spectral energy distribution. The pebble surface density is then taken to be the surface density in mm-sized particles at the peak of the profile. As such the values are uncertain due to several factors: (i) the simple surface density profile assumed by \citet{Andrews2011}; (ii) uncertainties in the underlying dust-particle distribution which could contribute to the mm-flux and (iii) the trap could be optically thick.  } 

Figure~\ref{fig:Sigma_crit_compare} shows that the vast majority of mm-bright ``transition'' discs have mm size particle surface densities sufficiently high that vortex formation due to low-mass planet formation is possible.  
 
Finally, it is well known that high viscosities can prevent vortex formation in the planet induced RWI mechanism and can also dissipate vortices \citep[e.g.][]{ValBorro2007,Fu2014b,ZhuStone14}. Our results are also consistent with the previous RWI results, in that large scale vortex formation is suppressed when the typical viscous alpha parameter is $>10^{-3}$. While typical values to explain the global evolution of the protoplanetary discs are of this order \citep[e.g.][]{Hartmann1998,Owen2011} there is every reason to expect the viscosity is likely to be lower in dust-traps and the outer regions of protoplanetary discs. Firstly, non-ideal MHD effects are important in the outer regions  of protoplanetary discs \citep[e.g.][]{Armitage2011} and in the case of ambipolar diffusion dominated discs, vortices are known to form and survive for an observable length of time \citep{ZhuStone14}. Secondly, the enhanced dust content in the dust trap is known to suppress the strength of MRI turbulence \citep[e.g.][]{Jacquet2012}. Therefore, it is not unreasonable to assume that the viscosity in the neighbourhood of the dust-trap is sufficiently low to allow the formation of vortices. 

\subsection{Long term evolution}

The long term evolution of the disc will be strongly controlled by whether it is able to form a vortex or not. As discussed above, if a large scale vortex is able to form it will trap all the pebbles in the vortex itself. As none of the vortices seen in the simulations are co-located with the orbiting planet \citep[see][for a discussion of how the planet and vortex may interact]{Ataiee2014}, then the vortex will starve the planet of pebbles and the rapid pebble accretion will cease. 

Our simulations suggested that vortex formation is fairly rapid $\sim 100$ orbits. This is similar to the dust trapping time-scale, thus we suspect that the pebbles will become easily trapped in the vortex. At this stage the pebbles will no longer be available to rapidly accrete onto the planet. If the planet's accretion source is quickly shut-off then it will still be luminous for a short period of time, as the gaseous envelope previously supported by the accretion contracts towards the planet. This contraction will maintain its luminosity for a time-scale roughly similar to the gaseous envelopes Kelvin-Helmholtz time-scale ($t_{\rm KH}$), which for a planet with an envelope mass considerably less than the solid mass is given by, \citep[e.g.][]{Lee2015}:
\begin{equation}
t_{\rm KH}=\frac{GM_p^2X_{\rm env}}{R_pL}=X_{\rm env}t_{\rm acc}
\end{equation}  

Thus, the Kelvin-Helmhotz time-scale is shorter than the accretion time-scale by the gas envelope to planet mass ratio. For these masses and luminosities $X_{\rm env}$ will be in the range $\sim$0.01, indicating that the luminosity output of the planet will drop on a time-scale $\sim 10-100$ orbits. This means that once the large scale vortex has trapped the majority of pebbles, the source of the barcolinicity generating the vortex will disappear on a comparable time-scale to the vortex formation time-scale, approximately as $R_T\propto t_{\rm KH}^{-1/2}$. Since a gravitationally contracting object will follow an evolution such that $t_{\rm KH}$ is approximately the time it has been cooling, then $R_T$ will decrease as the square root of time.   

Now, since the vortex lifetime is finite, as without a source of barcolinicty, viscosity, instabilities \citep[e.g.][]{Lesur2009} or even the inertia of the dust itself \citep[e.g.][]{Fu2014} can destroy a large scale vortex on a time-scale of roughly 1000s of orbits, which is short compared to the disc's lifetime (e.g. $> 10^4$ orbits).  We can therefore expect that vortex generation and dissipation occurs several times during the disc's lifetime, where pebble accretion onto a planet generates a vortex, which traps the pebbles hence suppressing the accretion and vortex generation. The vortex then dissipates after some time-scale, releasing the dust particles into an axi-symmetric ring that can then undergo pebble accretion onto a planet again, forming another vortex.  The entire process can repeat for a long time-scale until the pebble reservoir is too heavily depleted to permit vortex formation.  

If the pebble reservoir is too small to permit vortex formation, {\bc although as demonstrated by Equation~\ref{eqn:sigma_crit2} still significant to make the disc appear as a mm-bright disc.} As discussed in Section~2, pebble accretion onto a planet will then rapidly deplete its local reservoir on a time-scale $\lesssim 10^{5}$~years. {\bc This means the time-scale that a disc in this model would appear at intermediate mm-fluxes would be short.} Once the planet has depleted its local reservoir, particles at larger orbital separations will continue to drift into the dust trap. At large radii the maximum dust particle size is limited by drag rather than fragmentation \citep[e.g.][]{Birnstiel2012b}. Therefore, we can imagine that as dust particles drift from large radii in the disc, they will begin to grow and when they arrive in the vicinity of the planet they will be accreted readily as they will have naturally reached a size with $\tau_s=1$ at the planet's radius. In this stage, the planet's accretion rate will be limited by the accretion rate of dust particles into the trap due to drift, rather than the standard pebble accretion rate. Since these planets will necessary be low-mass ($< 10$~M$_\oplus$) they are unlikely to be able to accrete a significant gas envelope \citep[e.g.][]{Rafikov2006,Piso2014}. Thus, unlike the dust particles that could in principle be rapidly sequestered into low-mass planets, the gas will remain largely unchanged. This means this process may ultimately result in a low-mass dust-poor, gas-rich, low-mass-planet-rich disc, a type of disc for which there are poor observational constraints. 

Finally, the last concern is migration. These planets are low enough mass that they are unlikely to open a gap in the gas disc, where the gap opening mass is more typically $\sim 0.1$M$_J$ at tens of AU \citep[e.g.][]{Crida2006}. Therefore, migration is likely to take place in the type-I regime. While the migration rates for low-mass planets are still greatly uncertain in realistic protoplanetary discs, even the isothermal type-I migration rates, which are likely to strongly over-predict the migration rates, give rates $\sim 10^5$~years for a few earth mass planet at 10s of AU \citep[e.g.][]{Ward1997}. This means that a forming planet is able to initiate vortex formation and it will not migrate away in on the time scale for vortex formation to occur. However, migration might push the formed planet inside the transition disc cavity after the vortex has formed but before it has dissipated.  The vortex will certainly qualitatively effect the migration of such a low-mass planet (we note again we do not consider migration of our planet in our simulations presented here), as demonstrated by \citet{Ataiee2014}. This might mean that over a Myr lifetime of planet formation, the cycles of vortex generation and dissipation in the disc may dump a handful of low-mass planets inside the transition disc cavity, which could subsequently scatter.    

\subsection{Observational implications}

While we have suggested that transition discs are prime sites for planet formation through pebble accretion, the fact that it is so rapid means that if it was left to proceed as normal in discs with the parameters of standard mm-bright transition discs ($a\sim 10-50$~AU, $\Sigma_{\rm peb}\sim 1-10$g cm$^{-2}$), it would rapidly deplete the entire disc on an incredible short time $\lesssim 10^{5}$~years. We suggest the fact that these mm-bright transition discs can exist for a long enough time-scale to be observable, is that at the expected dust surface densities in these discs, the act of low-mass planet formation in their dust traps should result in large-scale vortex formation.

As discussed above these dust traps should trap all the pebbles within the vortex itself and prevent further planet growth. Therefore, many of these discs should spend {\bc some} fraction of their lifetimes with large scale vortices, which will be observable as large scale asymmetries when the discs are imaged at mm wavelengths. {\bcc For the mm-bright transition discs imaged at high resolution IRS48 \citet{vanderMarel2013} and HD142527 \cite{Casassus2013} show strong asymmetries, and LkH$\alpha$330 \citep{Isella2013} and SAO2016462 \citep{Perez2014} show weaker asymmetries. Many other observed transition discs do not show any evidence of an asymmetry (e.g. LkCa15, SR24S \citealt{vanderMarel2015}, Sz91 \citealt{Canovas2016}, DoAr 44 \citealt{vanderMarel2016}). In recent ALMA surveys of protoplanetary discs \citep[e.g.][]{Pascucci2016,Ansdell2016} several more transition disc like structures have been detected, many of which show no evidence for an asymmetry at $\sim 0.3"$ resolution. To date it is unclear exactly what fraction of transition discs show asymmetries. The reason for this is partly due to unsystematic observations at a variety of resolutions and partly due to the fluid definition of a ``transition disc'': for example should discs with large mm-holes but primordial SEDs be counted in this sample? \citep[See][for a discussion of these discs.]{Andrews2011,Owen2016}. What can be said with any confidence is that a moderate fraction of mm-bright transition discs show asymmetries.} Finally, exactly what kind of asymmetry our vortices produce will depend on the observational wavelength, surface density distribution and other factors (e.g. dust-growth and destruction) we do not consider here. Therefore, we must wait until dust-gas simulations are performed, before we can draw any hard conclusions from the (incomplete) transition disc statistics we have today. 

We suspect the hot-spot generated by the planet will fade on a time-scale of $\sim 100$ orbits. Thus if the vortex lifetime is significantly longer than 100 orbits, it would be very unlikely to directly observe the effect of the increase in local disc temperature in the continuum image as $R_T$ will have contracted to well inside the planet's Hill sphere once the vortex had trapped all the dust particles. However, it may be possible that the hot spot would generate some chemical fingerprint, that would last longer than the hot spot, spread into a ring, and possibly be a signature of this process.      

The critical pebble surface density for vortex formation derived in Equation~\ref{eqn:sigma_crit2} is similar to the value required to give a mm-flux at 140pc of $\sim 30$ mJy \citep[e.g.][]{Andrews2005}. This mm-flux value is the discriminating value between mm-bright and mm-faint transition discs determined by \citet{OC12}. \citet{OC12} and \citet{Owen2016} argued that mm-bright transition discs are likely to be rare and long lived (lifetimes $> 10^{5}-10^{6}$ years), whereas mm-faint transition discs are thought to be common (in the sense that all discs experience this phase) and short lived with lifetimes $\lesssim 10^{5}$~years. Here we suggest that this critical pebble surface density for vortex formation could provide the link as to why mm-bright transition discs are likely long lived and mm-faint transition discs are likely short lived. Since mm-bright transition discs can form vortices, they can prevent the dust from being rapidly turned into planets, producing many cycles of planet formation, vortex formation and subsequent destruction. 

Finally, one obvious consequence of this mechanism is the production of low-mass ($\lesssim$ Neptune mass) planets with orbital separations of tens of AU.  There is currently limited observational sensitivity to the low-mass exoplanet population at large separations.  However, mirco-lensing surveys have detected Neptune mass planets at separations $\sim$10~AU \citep[e.g.][]{Gaudi2012, Shvartzvald2016} and have suggested that such planets are common \citep[e.g.][]{Gould2006}.   The sensitivity of these experiments will only improve with future surveys hopefully revealing the full mass-spectrum and semi-major axis distribution of these systems.

\subsection{Future Directions}

In this work we have mainly argued that transition disc dust traps are likely to be prime sites of planet formation by pebble accretion. The rapid nature of the planetary accretion is likely to have many interesting consequences. Here we have argued that by modifying the temperature structure outside the gravitational influence of the planet it can provide a source of vorticity, which allows the growth of large scale vortices. 

In the simulations performed in this work we have described a very ideal setup, where we impose a local hot-spot that is not coupled to the subsequent dynamics. Specifically, we do not to attempt to model the coupled evolution of the dust, gas and planetary accretion in a self consistent way. While the time-scales suggest that large scale vortex formation is likely to occur the finer details of the results require further investigation. The vortex strength, how it traps dust particles, and when and how accretion onto the planet is shut-off due to the fact the pebbles are now trapped in the vortex requires coupled dust and gas simulations, possibly including grain growth. Such simulations will be able to investigate the cycle of planet formation, vortex growth, planet migration and dissipation that allows these discs to exist for the $>10^5-10^6$~years that we have postulated above.   

Furthermore, in-order to isolate the physics of the problem at hand we have neglected the self-gravity of the disc and the indirect potential that could effect the planetary dynamics and migration. In massive discs, vortex formation could result in the vortices transitioning to global ``fast'' modes \citep{Mittal2015}, similar to the transition of RWI generated vortices into fast modes discussed by \citep{Zhu16a,Zhu16b}. It maybe interesting to investigate the interaction between pebble accretion generated vortices and the indirect potential to see if fast modes can be triggered in this case, even without the original transition disc cavity.  

Finally, we note that while we have primarily focused on pebble accretion and vortex generation in transition discs, the mechanism of vorticity generation is not limited specifically to transition discs. Indeed, pebble accretion has been invoked to solve numerous planet formation problems at various locations and rates throughout primordial protoplanetary discs \citep[e.g.][]{Bitsch2015}. In our work, the gas cavity prevents the vortices from rapidly migrating allowing them to grow to large sizes.  In a primordial disc, we speculate that smaller vortices may be generated which can still trap particles and migrate. By trapping particles this could affect the assumed pebble accretion rates in such calculations. This may be particularly important at smaller radii, as our critical vortex formation threshold scales as $R^{-5/7}$, whereas disc surface densities are thought to fall in a steeper manner with radius, with mm observations of protoplanetary discs suggesting an $R^{-1}$ \citep{Andrews2009} decline, and the Minimum Mass Solar Nebula (MMSN) scaling as $R^{-3/2}$. Therefore, investigating the prospects of vortex generation in a primordial disc that is forming planets through pebble accretion is certainly a worthwhile investigation.   

\section{Summary}
We have argued that the dust-traps created within ``transition'' discs can serve as planet incubators, and consequently vortex generators.

 If a small $\gtrsim 10^{-4}$~M$_\oplus$ embryo forms, it would undergo rapid accretion through ``pebble accretion''.  The dust-trap naturally filters the dust particles size distribution within the trap,  such that most of the dust mass in the trap will be at the preferred size to undergo pebble accretion, namely those particles with Stokes numbers $\sim 1$. The high surface density of pebbles means accretion is extremely rapid, and the embyro can grow to masses $>1$~M$_\oplus$ on short time-scales ($10^4-10^5$~years). Thus, massive transition discs are prime site for the formation of low-mass $\lesssim 10$~M$_\oplus$ planets. Depending on the exact frequency of massive transition discs, planet formation in a transition disc induced dust trap could be a dominant mechanism of low-mass planet formation at large separation.  

Furthemore, we argue that the accretion luminosity liberated during the formation of the planet is large enough to heat the surrounding disc, well outside the planet's gravitational influence. This makes the disc {\it locally} baroclinic and unstable to vortex formation. By performing numerical simulations we show that these vortices will grow and merge until one large scale vortex is formed in about 100 orbits, only if the temperature of the disc can be increased by the planetary accretion outside the planet's Hill sphere. 

We suggest that this mechanism naturally explains the observed asymmetries in transition discs, as planet formation and rapid pebble accretion is difficult to prevent in the dust densities expected in transition disc dust trap and thus seemingly inevitable. Furthermore, our mechanism doesn't suffer from the requirement of a sharp density contrast required by the Rossby Wave Instability which may be difficult to generate in an actual protoplanetary disc. 

{\bc This new mechanism hinges on the production of a low-mass embryo to start undergoing pebble accretion (once the embryo forms it is difficult to imagine it could not undergo pebble accretion). We have not attempted to directly address the issue of the embryo's formation here. Several mechanisms do exist which would produces embryo's of the correct size (such as coagulation, the streaming instability of direct gravitational collapse, or some combination of them), they do require certain conditions be met. For example, the streaming stability requires low turbulence and high dust-to-gas ratios. The pressure traps to qualitatively provide many of these special conditions, it is impossible to say without quantitative calculations whether the observed ``transition'' discs satisfy these requirements all the time. }

Finally, we hypothesise a cycle of planet formation, vortex generation, dust-trapping and vortex dispersal where the duty cycle of vortex observability will be high.  Rapid planet formation will heat the disc and generate a large scale vortex. This large scale vortex will then trap the dust particles and prevent further pebble accretion and vortex growth. The vortex will then live for some time before being destroyed releasing all the dust particles in the axisymmetric dust-trap allowing the cycle to restart. Thus, our mechanism does not suffer from the problem that we require a long vortex lifetime for it to be observable, just a lifetime greater than a few hundred orbits.  Calculating this cycle end-to-end will require more detailed, probably coupled dust and gas simulations, however it presently offers a promising resolution to a number of outstanding observational and theoretical puzzles.

\section*{Acknowledgements}

The authors are grateful to the referee for advice that improved the manuscript. We are grateful to Richard Booth, Subo Dong, Ruobing Dong, Kaitlin Kratter, Tim Morton, Ruth Murray-Clay, Roman Rafikov, Giovanni Rosotti and Zhaohuan Zhu for interesting discussions. JEO acknowledges support by NASA through
Hubble Fellowship grant HST-HF2-51346.001-A awarded
by  the  Space  Telescope  Science  Institute,  which  is  operated  by  the  Association  of  Universities  for  Research
in  Astronomy,  Inc.,  for  NASA,  under  contract  NAS  5-26555.  JAK gratefully acknowledges support from the Institute for Advanced Study.

\label{sec:summary}





\begin{thebibliography}{999}

\bibitem[\protect\citeauthoryear{ALMA Partnership et al.}{2015}]{HLTau} ALMA Partnership, et al., 2015, ApJ, 808, L

\bibitem[\protect\citeauthoryear{Alexander \& Armitage}{2007}]{Alexander2007} Alexander R.~D., Armitage P.~J., 2007, MNRAS, 375, 500
\bibitem[\protect\citeauthoryear{Andrews \& Williams}{2005}]{Andrews2005} Andrews S.~M., Williams J.~P., 2005, ApJ, 631, 1134

\bibitem[\protect\citeauthoryear{Andrews et al.}{2009}]{Andrews2009} Andrews S.~M., Wilner D.~J., Hughes A.~M., Qi C., Dullemond C.~P., 2009, ApJ, 700, 1502 

\bibitem[\protect\citeauthoryear{Andrews et al.}{2011}]{Andrews2011} Andrews S.~M., Wilner D.~J., Espaillat C., Hughes A.~M., Dullemond C.~P., McClure M.~K., Qi C., Brown J.~M., 2011, ApJ, 732, 42 

\bibitem[\protect\citeauthoryear{Andrews et al.}{2016}]{Andrews2016} Andrews S.~M., et al., 2016, ApJ, 820, L40 


\bibitem[Ansdell et al.(2016)]{Ansdell2016} Ansdell, M., Williams, J.~P., van der Marel, N., et al.\ 2016, \apj, 828, 46 

\bibitem[\protect\citeauthoryear{Armitage}{2011}]{Armitage2011} Armitage P.~J., 2011, ARA\&A, 49, 195 



\bibitem[\protect\citeauthoryear{Ataiee et al.}{2013}]{Ataiee2013} Ataiee S., Pinilla P., Zsom A., Dullemond C.~P., Dominik C., Ghanbari J., 2013, A\&A, 553, L3 
\bibitem[\protect\citeauthoryear{Ataiee et al.}{2014}]{Ataiee2014} Ataiee S., Dullemond C.~P., Kley W., Reg{\'a}ly Z., Meheut H., 2014, A\&A, 572, A61 


\bibitem[\protect\citeauthoryear{Baruteau \& Zhu}{2015}]{Zhu16b} Baruteau C., Zhu Z., 2015, arXiv, arXiv:1511.03498

\bibitem[\protect\citeauthoryear{Birnstiel, Andrews, \& Ercolano}{2012}]{Birnstiel2012} Birnstiel T., Andrews S.~M., Ercolano B., 2012, A\&A, 544, A79 

\bibitem[\protect\citeauthoryear{Birnstiel, Klahr, \& Ercolano}{2012}]{Birnstiel2012b} Birnstiel T., Klahr H., Ercolano B., 2012, A\&A, 539, A148 


\bibitem[\protect\citeauthoryear{Birnstiel, Dullemond, \& Pinilla}{2013}]{Birnstiel2013} Birnstiel T., Dullemond C.~P., Pinilla P., 2013, A\&A, 550, L8 



\bibitem[\protect\citeauthoryear{Bitsch, Lambrechts, \& Johansen}{2015}]{Bitsch2015} Bitsch B., Lambrechts M., Johansen A., 2015, A\&A, 582, A112 


\bibitem[\protect\citeauthoryear{Calvet et al.}{2005}]{Calvet2005} Calvet N., et al., 2005, ApJ, 630, L185 

\bibitem[Canovas et al.(2016)]{Canovas2016} Canovas, H., Caceres, C., Schreiber, M.~R., et al.\ 2016, \mnras, 458, L29

\bibitem[Casassus et al.(2013)]{Casassus2013} Casassus, S., van der Plas, G., M, S.~P., et al.\ 2013, \nat, 493, 191

\bibitem[\protect\citeauthoryear{Casassus et al.}{2015}]{Casassus2015} Casassus S., et al., 2015, ApJ, 812, 126

\bibitem[\protect\citeauthoryear{Casassus}{2016}]{Casassus2016} Casassus S., 2016, PASA, 33, e013


\bibitem[\protect\citeauthoryear{Chiang \& Goldreich}{1997}]{Chiang1997} Chiang E.~I., Goldreich P., 1997, ApJ, 490, 368 

\bibitem[\protect\citeauthoryear{Clarke, Gendrin, \& Sotomayor}{2001}]{Clarke2001} Clarke C.~J., Gendrin A., Sotomayor M., 2001, MNRAS, 328, 485

\bibitem[\protect\citeauthoryear{Cieza et al.}{2007}]{Cieza2007} Cieza L., et al., 2007, ApJ, 667, 308 


\bibitem[\protect\citeauthoryear{Cieza et al.}{2008}]{Cieza2008} Cieza L.~A., Swift J.~J., Mathews G.~S., Williams J.~P., 2008, ApJ, 686, L115 

\bibitem[\protect\citeauthoryear{Crida, Morbidelli, \& Masset}{2006}]{Crida2006} Crida A., Morbidelli A., Masset F., 2006, Icar, 181, 587 
\bibitem[D'Alessio et al.(2001)]{dalessio2001} D'Alessio, P., Calvet, N., \& Hartmann, L.\ 2001, \apj, 553, 321 
\bibitem[\protect\citeauthoryear{Dullemond \& Dominik}{2005}]{Dullemond2005} Dullemond C.~P., Dominik C., 2005, A\&A, 434, 971 

\bibitem[Dr{\c a}{\.z}kowska et al.(2013)]{Drakowska2013} Dr{\c a}{\.z}kowska, J., Windmark, F., \& Dullemond, C.~P.\ 2013, \aap, 556, A37


\bibitem[Dr{\c a}{\.z}kowska \& Dullemond(2014)]{Drakowska2014} Dr{\c a}{\.z}kowska, J., \& Dullemond, C.~P.\ 2014, \aap, 572, A78

\bibitem[\protect\citeauthoryear{Espaillat et al.}{2010}]{Espaillat2010} Espaillat C., et al., 2010, ApJ, 717, 441 

\bibitem[\protect\citeauthoryear{Espaillat et al.}{2014}]{Espaillat2014} Espaillat C., et al., 2014, PPVI, 497 

\bibitem[Flock et al.(2015)]{Flock2015} Flock, M., Ruge, J.~P., Dzyurkevich, N., et al.\ 2015, \aap, 574, A68 



\bibitem[\protect\citeauthoryear{Fortney, Marley, \& Barnes}{2007}]{Fortney2007} Fortney J.~J., Marley M.~S., Barnes J.~W., 2007, ApJ, 659, 1661 

\bibitem[\protect\citeauthoryear{Fu et al.}{2014a}]{Fu2014b} Fu W., Li H., Lubow S., Li S., 2014, ApJ, 788, L41 

\bibitem[\protect\citeauthoryear{Fu et al.}{2014b}]{Fu2014} Fu W., Li H., Lubow S., Li S., Liang E., 2014, ApJ, 795, L39 

\bibitem[Gammie(1996)]{Gammie1996} Gammie, C.~F.\ 1996, \apj, 457, 355

\bibitem[\protect\citeauthoryear{Gaudi}{2012}]{Gaudi2012} Gaudi B.~S., 2012, ARA\&A, 50, 411 

\bibitem[Gorti et al.(2015)]{Gorti2015} Gorti, U., Hollenbach, D., \& Dullemond, C.~P.\ 2015, \apj, 804, 29 

\bibitem[\protect\citeauthoryear{Gould}{2006}]{Gould2006} Gould, A., Udalski, A., An, D., et al.\ 2006, \apjl, 644, L37 

\bibitem[\protect\citeauthoryear{Hammer, Kratter \& Lin}{2016}]{Hammer2016} Hammer M., Kratter K.~M., Lin, M.~K., 2016, MNRAS {\it in press}, arXiv:1610.01606

\bibitem[\protect\citeauthoryear{Hardy et al.}{2015}]{Hardy2015} Hardy A., et al., 2015, A\&A, 583, A66


\bibitem[\protect\citeauthoryear{Hartmann et al.}{1998}]{Hartmann1998} Hartmann L., Calvet N., Gullbring E., D'Alessio P., 1998, ApJ, 495, 385 

\bibitem[\protect\citeauthoryear{Haworth, Clarke, \& Owen}{2016}]{Haworth2016} Haworth T.~J., Clarke C.~J., Owen J.~E., 2016, MNRAS, 457, 1905 

\bibitem[Isella et al.(2013)]{Isella2013} Isella, A., P{\'e}rez, L.~M., Carpenter, J.~M., et al.\ 2013, \apj, 775, 30 

\bibitem[\protect\citeauthoryear{Jacquet \& Balbus}{2012}]{Jacquet2012} Jacquet E., Balbus S., 2012, MNRAS, 423, 437 


\bibitem[\protect\citeauthoryear{Johansen \& Youdin}{2007}]{Johansen2007} Johansen A., Youdin A., 2007, ApJ, 662, 627 

\bibitem[\protect\citeauthoryear{Johansen, Youdin, \& Mac Low}{2009}]{Johansen2009} Johansen A., Youdin A., Mac Low M.-M., 2009, ApJ, 704, L75 

\bibitem[\protect\citeauthoryear{Johansen \& Lacerda}{2010}]{Johansen2010} Johansen A., Lacerda P., 2010, MNRAS, 404, 475

\bibitem[\protect\citeauthoryear{Johansen et al.}{2015}]{Johansen2015} Johansen A., Mac Low M.-M., Lacerda P., Bizzarro M., 2015, SciA, 1, 1500109 


\bibitem[\protect\citeauthoryear{Kenyon \& Hartmann}{1987}]{Kenyon1987} Kenyon S.~J., Hartmann L., 1987, ApJ, 323, 714 

\bibitem[\protect\citeauthoryear{Kim et al.}{2009}]{Kim2009} Kim K.~H., et al., 2009, ApJ, 700, 1017 


\bibitem[Klahr \& Bodenheimer(2003)]{Klahr2003} Klahr, H.~H., \& Bodenheimer, P.\ 2003, \apj, 582, 869 

\bibitem[\protect\citeauthoryear{Koepferl et al.}{2013}]{Koepferl2013} Koepferl C.~M., Ercolano B., Dale J., Teixeira P.~S., Ratzka T., Spezzi L., 2013, MNRAS, 428, 3327 


\bibitem[\protect\citeauthoryear{Krauss et al.}{2007}]{Krauss2007} Krauss O., Wurm G., Mousis O., Petit J.-M., Horner J., Alibert Y., 2007, A\&A, 462, 977 

\bibitem[\protect\citeauthoryear{Lambrechts \& Johansen}{2012}]{Lambrechts2012} Lambrechts M., Johansen A., 2012, A\&A, 544, A32

\bibitem[\protect\citeauthoryear{Lee \& Chiang}{2015}]{Lee2015} Lee E.~J., Chiang E., 2015, ApJ, 811, 41

\bibitem[\protect\citeauthoryear{Lesur \& Papaloizou}{2009}]{Lesur2009} Lesur G., Papaloizou J.~C.~B., 2009, A\&A, 498, 1 


\bibitem[Lesur \& Papaloizou(2010)]{Lesur2010} Lesur, G., \& Papaloizou, J.~C.~B.\ 2010, \aap, 513, A60 

\bibitem[\protect\citeauthoryear{Li et al.}{2000}]{Li2000} Li H., Finn J.~M., Lovelace R.~V.~E., Colgate S.~A., 2000, ApJ, 533, 1023


\bibitem[\protect\citeauthoryear{Li et al.}{2001}]{Li2001} Li H., Colgate S.~A., Wendroff B., Liska R., 2001, ApJ, 551, 874

\bibitem[\protect\citeauthoryear{Lovelace et al.}{1999}]{Lovelace1999} Lovelace R.~V.~E., Li H., Colgate S.~A., Nelson A.~F., 1999, ApJ, 513, 805 

\bibitem[\protect\citeauthoryear{Lyra et al.}{2008}]{Lyra2008} Lyra W., Johansen A., Klahr H., Piskunov N., 2008, A\&A, 491, L41 


\bibitem[\protect\citeauthoryear{Lyra et al.}{2009}]{Lyra2009} Lyra W., Johansen A., Klahr H., Piskunov N., 2009, A\&A, 493, 1125 

\bibitem[\protect\citeauthoryear{Lyra \& Lin}{2013}]{LyraLin2013} Lyra W., Lin M.-K., 2013, ApJ, 775, 17

\bibitem[Lyra et al.(2015)]{Lyra2015} Lyra, W., Turner, N.~J., \& McNally, C.~P.\ 2015, \aap, 574, A10

\bibitem[\protect\citeauthoryear{Masset}{2000}]{Masset2000} Masset F., 2000, A\&AS, 141, 165 

\bibitem[\protect\citeauthoryear{van der Marel et al.}{2013}]{vanderMarel2013} van der Marel N., et al., 2013, Sci, 340, 1199

\bibitem[\protect\citeauthoryear{van der Marel et al.}{2015a}]{vanderMarel2015} van der Marel N., van Dishoeck E.~F., Bruderer S., P{\'e}rez L., Isella A., 2015, A\&A, 579, A106 

\bibitem[\protect\citeauthoryear{van der Marel et al.}{2015b}]{vanderMarel2015b} van der Marel N., Pinilla P., Tobin J., van Kempen T., Andrews S., Ricci L., Birnstiel T., 2015, ApJ, 810, L7 


\bibitem[\protect\citeauthoryear{van der Marel et al.}{2016}]{vanderMarel2016} van der Marel N., van Dishoeck E.~F., Bruderer S., Andrews S.~M., Pontoppidan K.~M., Herczeg G.~J., van Kempen T., Miotello A., 2016, A\&A, 585, A58 

\bibitem[\protect\citeauthoryear{Meheut et al.}{2012}]{Meheut2012} Meheut H., Meliani Z., Varniere P., Benz W., 2012, A\&A, 545, A134

\bibitem[\protect\citeauthoryear{Mer{\'{\i}}n et al.}{2010}]{Merin2010} Mer{\'{\i}}n B., et al., 2010, ApJ, 718, 1200-1223


\bibitem[\protect\citeauthoryear{Mittal \& Chiang}{2015}]{Mittal2015} Mittal T., Chiang E., 2015, ApJ, 798, L25 




\bibitem[\protect\citeauthoryear{Morbidelli et al.}{2015}]{Morbidelli2015} Morbidelli A., Lambrechts M., Jacobson S., Bitsch B., 2015, Icar, 258, 418 



\bibitem[\protect\citeauthoryear{Ormel \& Klahr}{2010}]{Ormel2010} Ormel C.~W., Klahr H.~H., 2010, A\&A, 520, A43



\bibitem[\protect\citeauthoryear{Owen, Ercolano, \& Clarke}{2011}]{Owen2011} Owen J.~E., Ercolano B., Clarke C.~J., 2011, MNRAS, 412, 13


\bibitem[\protect\citeauthoryear{Owen, Clarke, \& Ercolano}{2012}]{Owen2012} Owen J.~E., Clarke C.~J., Ercolano B., 2012, MNRAS, 422, 1880 



\bibitem[\protect\citeauthoryear{Owen \& Clarke}{2012}]{OC12} Owen J.~E., Clarke C.~J., 2012, MNRAS, 426, L96 

\bibitem[\protect\citeauthoryear{Owen et al.}{2013}]{Owen2013} Owen J.~E., Hudoba de Badyn M., Clarke C.~J., Robins L., 2013, MNRAS, 436, 1430 


\bibitem[\protect\citeauthoryear{Owen}{2014}]{Owen2014} Owen J.~E., 2014, ApJ, 789, 59 

\bibitem[\protect\citeauthoryear{Owen}{2016}]{Owen2016} Owen J.~E., 2016, PASA, 33, e005 



\bibitem[\protect\citeauthoryear{Padgett et al.}{2006}]{Padgett2006} Padgett D.~L., et al., 2006, ApJ, 645, 1283 

\bibitem[Pascucci et al.(2016)]{Pascucci2016} Pascucci, I., Testi, L., Herczeg, G.~J., et al.\ 2016, \apj, 831, 125 

\bibitem[\protect\citeauthoryear{P{\'e}rez et al.}{2014}]{Perez2014} P{\'e}rez L.~M., Isella A., Carpenter J.~M., Chandler C.~J., 2014, ApJ, 783, L13 

\bibitem[\protect\citeauthoryear{Piso \& Youdin}{2014}]{Piso2014} Piso A.-M.~A., Youdin A.~N., 2014, ApJ, 786, 21 




\bibitem[\protect\citeauthoryear{Pinilla, Benisty, \& Birnstiel}{2012}]{Pinilla2012} Pinilla P., Benisty M., Birnstiel T., 2012, A\&A, 545, A81 

\bibitem[\protect\citeauthoryear{Pinilla et al.}{2016}]{Pinilla2016} Pinilla P., Flock M., Juan Ovelar M., Birnstiel T.,  2016, A\&A, 596, A81 

\bibitem[Raettig et al.(2015)]{Raettig2015} Raettig, N., Klahr, H., \& Lyra, W.\ 2015, \apj, 804, 35 

\bibitem[Reg{\'a}ly et al.(2012)]{Regaly2012} Reg{\'a}ly, Z., Juh{\'a}sz, A., S{\'a}ndor, Z., \& Dullemond, C.~P.\ 2012, \mnras, 419, 1701 



\bibitem[\protect\citeauthoryear{Rice et al.}{2006}]{Rice2006} Rice W.~K.~M., Armitage P.~J., Wood K., Lodato G., 2006, MNRAS, 373, 1619


\bibitem[Ruge et al.(2016)]{Ruge2016} Ruge, J.~P., Flock, M., Wolf, S., et al.\ 2016, \aap, 590, A17 

\bibitem[\protect\citeauthoryear{Shvartzvald et al.}{2016}]{Shvartzvald2016} Shvartzvald, Y., Maoz, D., Udalski, A., et al.\ 2016, \mnras, 457, 4089 


\bibitem[\protect\citeauthoryear{Simon et al.}{2016}]{Simon2016} Simon J.~B., Armitage P.~J., Li R., Youdin A.~N., 2016, ApJ, 822, 55 



\bibitem[\protect\citeauthoryear{Strom et al.}{1989}]{Strom1989} Strom K.~M., Strom S.~E., Edwards S., Cabrit S., Skrutskie M.~F., 1989, AJ, 97, 1451 

\bibitem[\protect\citeauthoryear{Skrutskie et al.}{1990}]{Skrutskie1990} Skrutskie M.~F., Dutkevitch D., Strom S.~E., Edwards S., Strom K.~M., Shure M.~A., 1990, AJ, 99, 1187

\bibitem[\protect\citeauthoryear{Rafikov}{2006}]{Rafikov2006} Rafikov R.~R., 2006, ApJ, 648, 666 


\bibitem[\protect\citeauthoryear{de Val-Borro et al.}{2007}]{ValBorro2007} de Val-Borro M., Artymowicz P., D'Angelo G., Peplinski A., 2007, A\&A, 471, 1043 


\bibitem[\protect\citeauthoryear{Ward}{1997}]{Ward1997} Ward W.~R., 1997, Icar, 126, 261 


\bibitem[Windmark et al.(2012)]{Windmark2012a} Windmark, F., Birnstiel, T., G{\"u}ttler, C., et al.\ 2012, \aap, 540, A73 


\bibitem[Windmark et al.(2012)]{Windmark2012} Windmark, F., Birnstiel, T., Ormel, C.~W., \& Dullemond, C.~P.\ 2012, \aap, 544, L16 

\bibitem[\protect\citeauthoryear{Youdin \& Goodman}{2005}]{Youdin2005} Youdin A.~N., Goodman J., 2005, ApJ, 620, 459 

\bibitem[Zhang et al.(2014)]{Zhang2014} Zhang, K., Isella, A., Carpenter, J.~M., \& Blake, G.~A.\ 2014, \apj, 791, 42

\bibitem[\protect\citeauthoryear{Zhu et al.}{2011}]{Zhu2011} Zhu Z., Nelson R.~P., Hartmann L., Espaillat C., Calvet N., 2011, ApJ, 729, 47 


\bibitem[\protect\citeauthoryear{Zhu et al.}{2012}]{Zhu2012} Zhu Z., Nelson R.~P., Dong R., Espaillat C., Hartmann L., 2012, ApJ, 755, 6

\bibitem[\protect\citeauthoryear{Zhu et al.}{2014}]{Zhu2014} Zhu Z., Stone J.~M., Rafikov R.~R., Bai X.-n., 2014, ApJ, 785, 122

\bibitem[\protect\citeauthoryear{Zhu \& Stone}{2014}]{ZhuStone14} Zhu Z., Stone J.~M., 2014, ApJ, 795, 53 



\bibitem[\protect\citeauthoryear{Zhu \& Baruteau}{2015}]{Zhu16a} Zhu Z., Baruteau C., 2015, arXiv, arXiv:1511.03497 



\end{thebibliography}



\bsp	
\label{lastpage}
\end{document}